\documentclass[conference]{IEEEtran}
\IEEEoverridecommandlockouts

\usepackage{amsmath}
\usepackage{amssymb}
\usepackage{amsthm}

\usepackage{algorithm}
\usepackage{algpseudocode}

\usepackage{graphicx}
\usepackage{hyperref}
\usepackage{multirow}
\usepackage{subcaption}
\usepackage{threeparttable}
\usepackage{xcolor}
\usepackage{xspace}

\usepackage[normalem]{ulem}
\usepackage{enumitem}

\usepackage{etoolbox}

\usepackage{tikz}
\usetikzlibrary{plotmarks}
\usepackage{pgfplots}
\usetikzlibrary{patterns}
\pgfplotsset{compat=1.3}
\usetikzlibrary{backgrounds}
 
\newcommand{\circled}[2][]{%
	\tikz[baseline=(char.base)]{%
		\node[shape = circle, draw, fill=red, color=red, inner sep = .2pt]
		(char) {\phantom{\ifblank{#1}{#2}{#1}}};%
		\node at (char.center) {\makebox[0pt][c]{\color{white}{#2}}};}}
\robustify{\circled}

\newtheorem{definition}{Definition}
\newtheorem{theorem}{Theorem}
\newtheorem{lemma}{Lemma}

\usepackage{color} 
\usepackage{soul}
\usepackage{textcomp}
\usepackage{changes}
 
\soulregister\cite7
\soulregister\ref7
\soulregister\pageref7

\newcommand*{\myDots}{\ifmmode\mathellipsis\else.\kern-0.13em.\kern-0.13em.\fi} 

\newcommand{\nano}{\texttt{e-SeaFL}\xspace}
\newcommand{\APVC}{\mathtt{APVC}}

\newcommand{\cf}{{\it cf.}\xspace}
\newcommand{\eg}{{\it e.g.}\xspace}
\newcommand{\etal}{{\it et~al.}\xspace}

\newcommand{\ie}{{\it i.e.}\xspace}

\renewcommand{\vec}[1]{\mathbf{#1}}

\newcommand{\mySmallSkip}{\vskip 3pt} 

\newcommand\mal[1]{\textcolor{red}{#1}}
\newcommand\integrity[1]{\textcolor{blue}{#1}}

 \newcommand{\myPara}[1]{
    \mySmallSkip
    \noindent{\bf {#1}{.}}
 }

\newcommand{\randasgn}{\stackrel{\smash{\$}}{\leftarrow}}

\newcommand{\sk}{\mathsf{sk}}
\newcommand{\pk}{\mathsf{pk}}

\newcommand{\Kg}{\mathsf{Kg}}
\newcommand{\Sgn}{\mathsf{Sgn}}
\newcommand{\Vfy}{\mathsf{Vfy}}
\newcommand{\Hash}{\mathsf{Hash}}
\newcommand{\PRG}{\mathsf{PRG}}
\newcommand{\PRF}{\mathsf{PRF}}

\newcommand{\AEnc}{\mathsf{AEnc}}
\newcommand{\ADec}{\mathsf{ADec}}
\newcommand{\SEnc}{\mathsf{SEnc}}
\newcommand{\SDec}{\mathsf{SDec}}

\newcommand{\Mu}{\mathsf{mu}}
\newcommand{\Ex}{\mathsf{ex}}
\newcommand{\Su}{\mathsf{su}}

\newcommand{\Ka}{\mathsf{Ka}}
\newcommand{\PIL}{\mathsf{PI}_\mathsf{L}}

\newcommand{\aeenc}{\mathtt{AE}.\mathtt{enc}}
\newcommand{\aedec}{\mathtt{AE}.\mathtt{dec}}

\newcommand{\Rssh}{\mathsf{Rssh}}
\newcommand{\Rsshe}{\mathsf{Rssh}_\mathsf{e}}
\newcommand{\sshl}{\mathsf{ssh}_\mathsf{l}}

\newcommand{\sgnkeygen}{\mathtt{Kg}}
\newcommand{\sgnsign}{\mathtt{Sgn}}
\newcommand{\sgnverify}{\mathtt{Vfy}}

\newcommand{\kakeygen}{{\mathtt{Kg}}}
\newcommand{\kaagree}{{\mathtt{Ka}}}

\newcommand{\FlamingoServerFirst}{Flamingo$_{64}$}
\newcommand{\FlamingoServerSec}{Flamingo$_{128}$}
\newcommand{\FlamingoUserFirst}{Flamingo$_{64}$(R)}
\newcommand{\FlamingoUserSec}{Flamingo$_{128}$(R)}
\newcommand{\FlamingoDecFirst}{Flamingo$_{64}$(w/D)}
\newcommand{\FlamingoDecSec}{Flamingo$_{128}$(w/D)}

\newcommand{\node}{\mathtt{A}}

\newcommand{\agg}{\mathtt{S}}

\newcommand{\adv}{\mathtt{Adv}}

\newcommand{\user}{\mathtt{P}}

\newcommand{\calP}{\mathcal{P}}
\newcommand{\calA}{\mathcal{A}}

\newcommand{\x}{{\mathtt{x}}}

\newcommand{\mk}{\mathtt{mk}}

\newcommand{\VC}{\mathtt{VC}}
\newcommand{\Setup}{\mathtt{Setup}}
\newcommand{\Comm}{\mathtt{Comm}}
\newcommand{\KeyGen}{\mathtt{KeyGen}}

\newcommand{\cm}{\mathtt{cm}}
\newcommand{\pp}{\mathtt{pp}}
\newcommand{\negl}{\mathtt{negl}}

\newcommand{\secParam}{\kappa}
\newcommand{\getsRandom}{\stackrel{\$}{\gets}}

\newcommand{\set}[1]{\mathcal{#1}}
\newcommand{\Ulist}{\mathcal{L}}

\newcommand{\Sim}{\mathtt{Sim}}

\newcommand{\real}{\mathtt{REAL}}
\newcommand{\simul}{\mathtt{Simul}}

\newcommand{\linebreakand}{%
  \end{@IEEEauthorhalign}
  \hfill\mbox{}\par
  \mbox{}\hfill\begin{@IEEEauthorhalign}
}

 \newcommand{\textasciitilden}{\raisebox{0.5ex}{\texttildelow}}

\begin{document}

\title{Efficient Secure Aggregation for Privacy-Preserving Federated Machine Learning
}

\author{
  \IEEEauthorblockN{Rouzbeh Behnia}
  \IEEEauthorblockA{
    \textit{University of South Florida} \\
    behnia@usf.edu}\\
  \IEEEauthorblockN{Sherman S. M. Chow}
  \IEEEauthorblockA{
    \textit{The Chinese University of Hong Kong} \\
    smchow@ie.cuhk.edu.hk} 
   \and
   \IEEEauthorblockN{Arman Riasi}
  \IEEEauthorblockA{
    \textit{Virginia Tech} \\
    armanriasi@vt.edu}\\
  \IEEEauthorblockN{Balaji Padmanabhan}
  \IEEEauthorblockA{
    \textit{University of Maryland, College Park} \\
    bpadmana@umd.edu}
  \and
  \IEEEauthorblockN{Reza Ebrahimi}
  \IEEEauthorblockA{
    \textit{University of South Florida} \\
    ebrahimim@usf.edu}\\
  \IEEEauthorblockN{Thang Hoang}
  \IEEEauthorblockA{
    \textit{Virginia Tech} \\
    thanghoang@vt.edu}
}

\maketitle
\pagestyle{plain} 

\begin{abstract}
Secure aggregation protocols ensure the privacy of users' data in federated learning by preventing the disclosure of local gradients.
Many existing protocols impose significant communication and computational burdens on participants and may not efficiently handle the large update vectors typical of machine learning models.
Correspondingly, we present \nano, an efficient verifiable secure aggregation protocol taking only one communication round during the aggregation phase. 
\nano allows the aggregation server to generate proof of honest aggregation to participants via authenticated homomorphic vector commitments.
Our core idea is the use of assisting nodes to help the aggregation server, under similar trust assumptions existing works place upon the participating users. 
Our experiments show that the user enjoys an order of magnitude efficiency improvement over the state-of-the-art (IEEE S\&P 2023)  for large gradient vectors with thousands of parameters. 
Our open-source implementation is available at 
\url{https://github.com/vt-asaplab/e-SeaFL}. 
\end{abstract}






\section{Introduction}
Federated learning (FL) enables the training of centralized machine learning (ML) models through the \mbox{contribution} of distributed parties while preserving the privacy of their local data~\cite{aistats/McMahanMRHA17,corr/KonecnyMRR16}.
Instead of sharing raw data, each participant trains a ``local model'' on their local dataset.
After each iteration, the local gradient updates---reflecting the adjustments needed to improve the global model---are sent back to a central server, which then aggregates these gradients to refine the global model by computing their average.

Unfortunately, local gradients can inadvertently expose sensitive information~\cite{ccs/HitajAP17} about user data.
Observing that the server only requires the element-wise weighted average of the gradient vectors for model updates rather than access to individual gradients,
we can utilize secure aggregation protocols~\cite{eurocrypt/NaorPR99,ccs/ChaseC09,ccs/BonawitzIKMMPRS17} to compute the sum of the gradients (followed by a division of a publicly known number of participants).

Cryptographic mechanisms underlying a secure aggregation protocol
keep individual inputs (\ie, gradients here) private.
The protocol of Bonawitz~\etal~\cite{ccs/BonawitzIKMMPRS17} lets each user mask their gradients with pseudorandom terms symmetrically shared between user pairs and then secret-shared among other users~\cite{eurocrypt/NaorPR99,ccs/ChaseC09}.
They~\cite{ccs/BonawitzIKMMPRS17} also outlined the core practical challenges of such protocols, including:
\begin{enumerate}
\item
Scalable processing of high-dimensional gradients, say foundation models with billions of parameters~\cite{corr/abs-2108-07258,mima/FloridiC20}.
\item
Security against malicious attacks in a server-mediated environment.
\item
Minimizing the setup complexity for users with constrained devices and unreliable connections.
\item
Correcting computation affected by user dropouts.
\end{enumerate}

Kicking off a wave of innovation, many manifestations of multi-party computation (MPC) have been developed.
Most require multiple communication rounds for aggregation with inbound and outbound communication for $n$ users summarized in Table~\ref{tab:usercomp}, 
including 
BIK$^+$17~\cite{ccs/BonawitzIKMMPRS17},
BBG$^+$20~\cite{ccs/BellBGL020},
MicroSecAgg~\cite{popets/GuoPSBB24}\footnote{
MicroSecAgg$_{\text{gDL}}$~\cite[\S4.1]{popets/GuoPSBB24} and MicroSecAgg$_{\text{CL}}$~\cite[\S5]{popets/GuoPSBB24} employ user grouping (\cf~\cite{ccs/BellBGL020}).
MicroSecAgg$_{\text{DL}}$~\cite[\S4.5]{popets/GuoPSBB24} 
uses a
single group.
},
and Flamingo~\cite{sp/MaWAPR23}.
Other works use homomorphic encryption (HE), \eg,~\cite{usenix/ZhangLX00020, ccs/DasuSM22},
which is round-efficient but often incurs huge computational overhead for large models.
Typically, MPC-based methods only tolerate a small fraction of user dropouts, as enough shares are needed to reconstruct the secrets of the dropout users.
Honest aggregation of servers is also assumed.
Errors or malicious behavior at the server~\cite{cvprw/CaoG22,uss/FangCJG20,icml/BhagojiCMC19} remain undetectable.
Results of such deviations could be catastrophic (\eg,~\cite{ccs/MilajerdiEGV19}).

We propose \nano, a single-round secure aggregation protocol that supports multiple training tasks with a one-time setup.
For a single-element update, 
our communication cost is $O(k)$, where $k$ is the number of assisting nodes and can be as small as~$2$.
Notably, for gradient vectors with $16,000$ weights and $1,000$ users (Figure~\ref{fig:exp:IM}), 
\nano outperforms state-of-the-art protocols Flamingo~\cite{sp/MaWAPR23} and MicroSecAgg~\cite{popets/GuoPSBB24}
by up to two and five orders of magnitude, 
respectively.
\nano scales out to large numbers of users and is highly robust to user dropouts.
Additionally, {\nano} introduces a proof-of-honest-aggregation protocol that ensures model integrity against a malicious server, addressing a known challenge in verifying the computational correctness of randomly-masked inputs\mbox{~\cite{tifs/XuLL0L20,iacr/Guo22,tdsc/HahnKKH23}}.
\nano is secure in both the semi-honest and malicious settings.

\subsection{Technical Highlights}
To hide local gradient updates $w_i$ using the sum of pseudorandom values, each user $\user_i$ derives a masking term~$h_i$, ensuring that these terms cancel out when the updates are aggregated, \ie, $\sum_{i \in [n]} h_i = 0$.
To achieve this, every pair of users $\user_i$ and $\user_j$ employs a (non-interactive) key agreement protocol to derive a shared key $\mk_{i,j}$.
This key is then used to generate masks for each iteration via a pseudorandom function (PRF) family; the masks should be ``symmetric'' to ensure they cancel when both $\user_i$ and $\user_j$ contribute their updates.
To account for user dropouts, $\mk_{i,j}$ is secret-shared using, for example, a $t$-out-of-$n$ Shamir secret-sharing scheme~\cite{cacm/Shamir79}, providing robustness.

\begin{table}[t!]
	\caption{User communication in every round for a training iteration of a single-element update with malicious security} \label{tab:usercomp}
		\begin{tabular}{| c c c c c c |}
			\hline
			\textbf{Rd.} & 	
   \cite{ccs/BonawitzIKMMPRS17} & 
   \cite{ccs/BellBGL020}\! & 
   \cite[\S4.5]{popets/GuoPSBB24}\! & 
   \cite[\S4.1\&5]{popets/GuoPSBB24}
	& \cite{sp/MaWAPR23}
	\\
	\hline \hline
			
$1$ & $O(n)$ & $O(1)$\! & $O(1)$\! & $O(1)$\! & $O(n)$
\\ \hline 
$2$ & $O(n)$ & $O(\log n)$\! & $O(n)$\! &$O(\log n)$\! & $O(\log n)$
\\ \hline 
$3$ & $O(1)$ & $O(\log n)$\! & $O(1)$\! & $O(1)$\! & $O(n\log n)$
\\ \hline
$4$ & $O(n)$ & $O(1)$\! & & &
\\ \hline 
$5$ & $O(n)$ & $O(\log n)$\! & & &
\\ \hline 
$6$ & & $O(\log n)$\! & & &
\\ \hline 
			\end{tabular}
\end{table}

\noindent \textbf{1.~Assisting nodes based on similar trust assumptions.}
Most existing works (\eg,~\cite{popets/GuoPSBB24,sp/MaWAPR23,ccs/BonawitzIKMMPRS17,popets/GuoPSBB24}) \emph{assume} that a subset of users remains honest.
On one hand, increasing data contributions from more users is desirable.
On the other hand, a larger user base also increases the likelihood of having more dishonest or compromised users, potentially outnumbering honest ones.

To address this, we introduce a set of \emph{assisting nodes}, each of which shares a secret seed with a user and aids the server in unmasking the final model by providing partial aggregated masking terms for participating users.
Similar \emph{assisting} entities have been used in other domains (\eg,\mbox{~\cite{ccs/DavidM0NT22,iotj/YuCZGW21,acsac/BehniaY21,infocom/WangTSLM0C18}}) and have also been applied in the state-of-the-art aggregation protocol~\cite{sp/MaWAPR23}.

Our security guarantee ensures that a user's local gradients remain secure even if all but one of the assisting nodes is compromised.
This design enables more efficient setup and aggregation phases.

A rotating set of users can act as assisting nodes during each training iteration, as the overhead on assisting nodes is comparable to, or even lower than, that imposed on users in the state-of-the-art~\cite{popets/GuoPSBB24,sp/MaWAPR23} (see Figure~\ref{fig:exp:IM}).
To ensure resiliency against assisting nodes going offline, we can utilize secret sharing~\cite{cacm/Shamir79} or key-homomorphic PRF \cite{crypto/BonehLMR13} to reconstruct the offline nodes secret or masking terms (Section~\ref{sec:Resiliency}) without adding extra burden on users.

\mySmallSkip
\noindent
\textbf{2.~Proof of honest aggregation.}
The security models of existing works~\cite{ccs/BonawitzIKMMPRS17,ccs/BellBGL020,sp/MaWAPR23,popets/GuoPSBB24} primarily address privacy attacks but fail to ensure the integrity of the final model computed by the aggregation server in each iteration.
Previous efforts to provide proof of integrity have either been impractical~\cite{tdsc/HahnKKH23,tifs/XuLL0L20} or proven insecure (\eg,~\cite{tifs/GuoLLGHDB21,iacr/Guo22}).

\nano uses authenticated homomorphic vector commitment~\cite{ndss/LeHYSH23} to provide proof of honest aggregation, specifically targeting threats posed by a malicious server.\footnote{
Mitigating threats from malicious clients would require third-party access to training data, which could undermine the data privacy guarantees of secure aggregation protocols (see Section~\ref{sec:threatmodelintegrity}).
}
Each user computes a commitment on their gradient and the homomorphism allows users to verify the integrity of the final aggregated model.

\subsection{Related Work} 
Several distinct research directions have been explored.
Heterogeneous environments, where users possess varying computational and communication capacities, are also considered~\cite{corr/abs-2112-12872, tcom/ElkordyA22}, with linear overheads comparable to earlier works~\cite{ccs/BonawitzIKMMPRS17, ccs/BellBGL020}.
Extensions for a differentially private protocol based on the learning-with-errors assumption~\cite{uss/StevensSVRCN22} have also been introduced, maintaining similar overheads~\cite{ccs/BellBGL020}.
Random graph topologies have also been leveraged to reduce communication overhead~\cite{corr/abs-2012-05433}.
Multi-key HE can encrypt the gradients so that they can be aggregated without decryption, which requires the (distributed) secret key(s)~\cite{ijis/MaNSL22}.
A generic semi-honest-secure construction from MPC or HE has also been proposed~\cite{sp/FereidooniMMMMN21}, without reporting the performance metrics for the HE instantiation.
New ideas are needed to enhance communication efficiency.

Two notable recent works, MicroSecAgg\footnote{
It was first published online as MicroFedML on \href{https://eprint.iacr.org/archive/2022/714/20220605:035426}{June 5, 2022}, which claimed Flamingo~\cite{sp/MaWAPR23} to be their subsequent work.
}
and Flamingo offer different approaches to tackling communication efficiency and provide a one-time setup phase.
Flamingo~\cite{sp/MaWAPR23}, a recent development based on MPC and linear HE, introduces assisting nodes called decryptors. 
The main idea is to utilize threshold decryption \cite{crypto/DesmedtF89} to encrypt the shared and individual masking terms, following the approach of \cite{ccs/BonawitzIKMMPRS17,ccs/BellBGL020}, and attach them to the updates submitted by the user.
This requires three rounds of communication to perform secure aggregation: The masked input vectors and ciphertexts are submitted to the server in the first round. During the second round, the clients' online/offline status is verified with the help of decryptors. In the final round, the required information to compute the unmasked sum is provided to the server. 
Flamingo requires a $2/3$ threshold of honest decryptors, which, according to established bounds~\cite{sp/MaWAPR23}, necessitates at least $64$ decryptors being active during each iteration. 

MicroSecAgg~\cite{popets/GuoPSBB24} introduces $3$-round single-server protocols.
One of which is especially computationally efficient 
as it is tailored to aggregation in the exponent, assuming a pre-computed discrete logarithm table.
This restriction remains relevant in various scenarios, including survey responses,
voting, 
browser telemetry, and
contact tracing, 
or gradient sparsification, 
quantization, and 
weight regularization
in FL.
MicroSecAgg excels for a large number of users but suffers for high-dimensional weight/input vectors as the number of exponentiations 
scales with it.
Flamingo does not suffer from this issue.
Both Flamingo and MicroSecAgg assume a static adversary, as the neighborhoods for each iteration are predetermined during the initial setup phase.
\section{Preliminaries}\label{sec:prelim} 

\noindent \textbf{Notations.}
Vectors are denoted by bold lowercase letters, \eg,~$\Vec{a}$.
$a \randasgn \mathbb{Z}$ denotes $a$ is randomly sampled from $\mathbb{Z}$.
Let $\PRF: K \times M \rightarrow Y$ be a pseudorandom function indexed by a key $\sk \in K$, which maps an input $m \in M$ to an output $y \in Y$.
$[x]$ denotes the set $\{1, \dots, x\}$.
Let $\vec{g} = (g_1, \ldots, g_d)$ and $\vec{x} = (x_1, \ldots, x_n)$,
we denote $\vec{g}^{\vec{x}} = \prod_{i = 1}^d g_i^{x_i}$.
Due to space limits, cryptographic notations are deferred to Appendix~\ref{sec:cryptobg}.


 \setlength{\textfloatsep}{0.1cm}

\begin{figure}
	\centering
	\includegraphics[width = 0.4\textwidth]{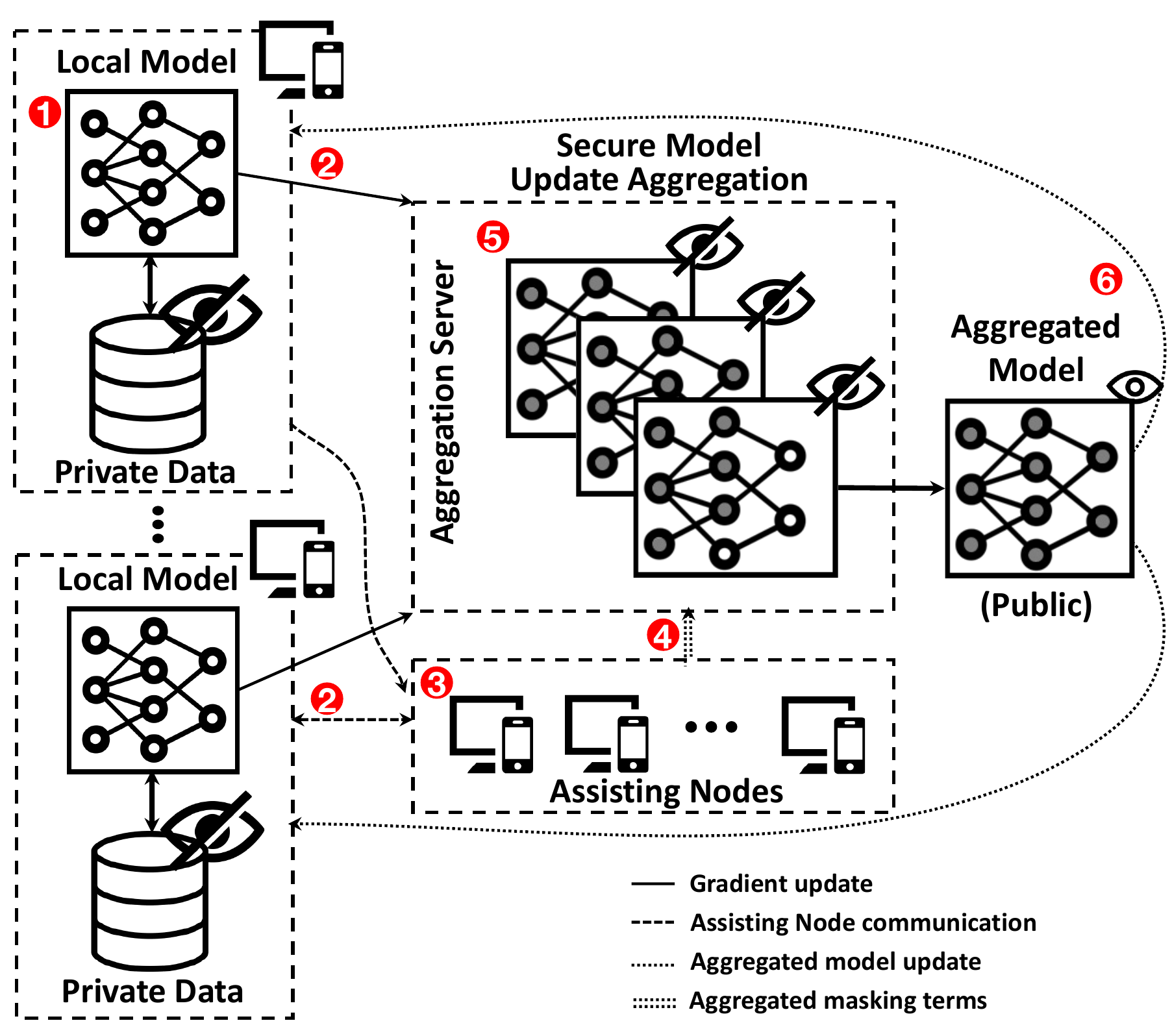}
	\caption{System model of \nano}
	\label{fig:sysmodel}
\end{figure}
\setlength{\floatsep}{0.1cm} 

For model integrity proof, we employ an authenticated Pederson vector commitment (APVC)~\cite{ndss/LeHYSH23}.
APVC permits the commitor to commit to a vector with keyed verification, requiring an extra key generation algorithm.

\begin{definition}[APVC]\label{def:apvc}
Given $\set{M}, \set{R} = \mathbb{Z}_p$, $\set{C} = \mathbb{G}$ of order $p$, an authenticated Pedersen vector commitment $\APVC:(\Setup,\KeyGen,\Comm)$ is defined as follows.

	\begin{itemize}
		 \item $ (\Vec{g},h)\gets \Setup(1^\kappa,d)$: Takes as input a security parameter $\kappa $ and an integer $d$ it outputs a vector $\vec{g} = (g_1, \dots, g_d)$ and $ h \getsRandom \mathbb{G}$.
	\item $\rho \gets \KeyGen(1^\kappa)$:
	Takes as input a security parameter $\kappa $ and outputs $\rho \getsRandom \mathbb{Z}_p$.
	\item $\cm \gets \Comm(\rho,\vec{x},r):$ Takes as input an input vector $\vec{x} = (x_1, \dots, x_d)$, $\rho$ and $r$ and outputs the commitment $\cm\gets h^r\vec{g}^{\rho \vec{x}} = h^r \prod_{i = 1}^n g_i^{\rho x_i} \in \mathbb{G}$.
	\end{itemize}
\end{definition}

$\APVC$ is homomorphic under $(\mathbb{Z}_p^n,+)$, $(\mathbb{Z}_p,+)$, and $(\mathbb{G}, \cdot)$.
Specifically,
$\Comm(\rho,\vec{x}_1,r_1) \cdot \Comm(\rho,\vec{x}_2,r_2) = \allowbreak
\Comm(\rho, \vec{x}_1 + \vec{x}_2, r_1+ r_2)$.

\subsection{System Model } \label{sec:sysmodel}
\begin{definition}[\cite{ccs/BonawitzIKMMPRS17,ccs/BellBGL020}]\label{def:aggprotc}
An aggregation protocol $\Gamma$,
with a set of users $\mathcal{P}$, a set of assisting nodes $\mathcal{A}$, 
an aggregation server~$\agg$,
and an integer $T$, takes place in two phases:

\begin{itemize}

\item \emph{Setup}:
A one-time phase mainly to set up the keys.

\item \emph{Aggregation}: 
Each party $\user_i$ computes a local update $w_t^{\user_i}$ to be sent to $\agg$, which outputs the update 
$w_t = \sum_{\user_i \in \mathcal{P}}w_t^{\user_i}$.
This phase runs for $T$ iterations.
\end{itemize}

Correctness with dropout rate $\delta$ is said to hold if,
for each iteration $t \in [T]$, 
for all the sets of offline users 
$\mathcal{P}_o \subset \mathcal{P}$,
where 
$|\mathcal{P}_o| < \delta|\mathcal{P}|$,
the server outputs 
$w_t = \sum_{\user_i \in \mathcal{P} \setminus \mathcal{P}_o}{w_t^{\user_i}}$
with an overwhelming probability.
\end{definition}

As Figure~\ref{fig:sysmodel} illustrates,
our system consists of:
\mySmallSkip
\noindent
1) $n$ users (parties)
$\{\user_1, \dots, \user_n\}$, who collaboratively train a shared model by computing local gradient vectors $\Vec{w}^{\user_i}_{t}$ on their private data during each iteration $t$.
\mySmallSkip
\noindent
2) A set of $k$ assisting nodes $\{\node_1, \dots, \node_k\}$, which assist server $\agg$ in aggregating the local updates.
Rotating users can take on the role of assisting nodes in each iteration.
\mySmallSkip
\noindent
3) A central aggregation server $\agg$, which is in charge of aggregating the locate updates $\Vec{w}_{t}^{\user_i}$ of users and sending the updated model $\Vec{w}_t$ back to them after each iteration.

On top of existing system models~\cite{ccs/BonawitzIKMMPRS17,ccs/BellBGL020,popets/GuoPSBB24},
we introduce $k$ assisting nodes.
They only communicate with the user during the initialization.
After which, they only receive
messages from the users.

Below is the workflow in each phase.
 
\noindent \emph{Setup phase}:
This is a one-time process for
key generation, 
and is supposed to conclude in one round.
In the end, 
users $\{\user_1, \dots, \user_n\}$ and assisting nodes $\{\node_1, \dots, \node_k\}$ are 
equipped with the necessary secret keys, including $\x$ (from key exchanges and possibly local key generation).
In the malicious-secure setting, 
users, assisting nodes, and the aggregation server also receive a copy of all user public keys.

\smallskip
\noindent \emph{Aggregation phase}:
This phase consists of two rounds:
\emph{Masking Updates} and \emph{Aggregate Updates}.
As in Figure~\ref{fig:sysmodel}:
\mySmallSkip
Step~\circled[1]{\small 1}:
Each party $\user_i$, after computing the local update $\vec{w}^{\user_i}_t$, computes the masked update $\vec{y}_t^{\user_i}$.
%
\mySmallSkip
Step~\circled[1]{\small 2}:
$\user_i$ sends the masked update $\vec{y}_t^{\user_i}$ and a participation message (with the iteration number $t$) to the aggregation server $\agg$ and the $k$ assisting nodes, respectively.
\mySmallSkip
Step \circled[1]{\small 3}:
Upon participation message $t$ from $\user_i$,
each assisting node $\node_j$ computes an aggregated masking vector~$\vec{a}^{\node_j}_t$.
\mySmallSkip
Step~\circled[1]{\small 4}:
All of $\{\node_1, \dots, \node_k\}$ send their aggregated masks $\vec{a}^{\node_j}_t$ (with signatures) to $\agg$ (for malicious security).
\mySmallSkip
Step \circled[1]{\small 5}:
{$\agg$ aggregates the final model using the users' local masked updates $\vec{y}_t^{\user_i}, \forall i \in [n]$ and $\{\vec{a}^{\node_j}_t$\}.
\mySmallSkip
Step \circled[1]{\small 6}:
$\agg$ replies with the updated model $\vec{w}_t$.
 
For model integrity against the malicious server, 
in the setup phase, one assisting node also generates a system-wide authenticated secret and disseminates it to all parties.\footnote{
To avoid a single point of failure, the protocol can be initiated by having all (or a subset of) assisting nodes generate different system-wide authenticated secrets with a small cost in communication and computation.
}
In the aggregation phase, after computing local updates, each user needs to send an authenticated commitment to the aggregation server.
In the final step of the Aggregate updates round, $\agg$ disseminates proof of honest aggregation, allowing users to verify if the final model was genuinely constructed by aggregating the local updates.

\subsection{Threat Models} \label{sec:ccs/MilajerdiEGV19model}

\subsubsection{Privacy}
Our adversary can control the aggregation server and a bounded fraction of the users and assisting nodes.\footnote{
Inference attacks after the model is established and deployed, 
\eg, membership inference via repeated queries, 
are beyond the scope of secure-aggregation-based training, 
and can be addressed by differential privacy.
}
Its goal is to learn any information about the individual updates $\Vec{w}^{\user_i}_t$.
To capture the unstable network situations, the adversary can temporarily drop or permanently disconnect a $\delta$ fraction of users.
We only assume one of the assisting nodes is honest for the privacy of the user updates.

In the semi-honest setting, parties corrupted by the adversary run the protocol honestly but attempt to infer secret information by observing the messages they receive from other parties, 
\eg, a group of passive but colluding parties pooling their views to learn about the secret information.

A malicious adversary possesses all the capabilities of semi-honest adversaries but can also deviate from the protocol, \eg, 
sending wrong messages or not sending messages.

Following~\cite{ccs/BellBGL020}, we define a summation protocol as being $\alpha$-secure if it can guarantee that each honest party's input is aggregated with the minimum of $\alpha|\calP_H|$ other secrets, where $\calP_H$ is the set of honest parties.
Similar to~\cite{ccs/BellBGL020}, we use this summation protocol in our simulation-based proof.

\begin{definition}
[$\alpha$-summation ideal functionality]
\label{def:alphasummation}
For integers $p,n,d$ and $\alpha \in [0,1]$, let $L\subseteq [n]$ and $\mathcal{X}_L: = \{\Vec{x}_i\}_{i\in {L}}$ where $\Vec{x}_i \in\mathbb{Z}_p^d$.
For $Q_{L}$ as the set of partitions of $L$ and a collection of pairwise disjoint subsets $ \{L_{1}, \dots, L_l\} \in Q_{L}$, the $\alpha$-summation ideal functionality for all subsets in $Q_{L}$ is denoted as
$\mathcal{F}_{\vec{x},\alpha}(\cdot)$ and computes $\mathcal{F}_{\vec{x},\alpha}(\{L_i\}_{i\in[1, \dots, l]}) \rightarrow \{\Vec{s}_i\}_{i\in[1, \dots, l]}$
where 
\begin{equation*}
\forall j\in[1, \dots, l] ,\Vec{s}_{j} = 
	\begin{cases}
	\sum_{j\in Q_L}\vec{x}_j & \text{if}~Q_L|\geq \alpha|L| \\
	
	\bot & \text{otherwise.}
	\end{cases} 
\end{equation*}

\end{definition}


Following~\cite{ccs/BonawitzIKMMPRS17,popets/GuoPSBB24}, we define the notion of privacy against semi-honest/malicious adversaries in the following.

\begin{definition}[Privacy of secure aggregation protocols]\label{def:aggSec}
Given a secure signature scheme $\Pi$ and a key exchange protocol $\Sigma$ instantiated with security parameter $\kappa$, 
an aggregation protocol $\Gamma$ parameterized by $(\calP, \calA, \agg, T)$, as defined in Definition~\ref{def:aggprotc}, provides privacy against a semi-honest (malicious) adversary $\adv$ if, there is exist a PPT simulator $\mathtt{Sim}$ that for any iteration $t \in [T]$, and all input vectors $\mathcal{X}^t = \{\Vec{x}_1, \ldots, \Vec{x}_n\}$, generates an output that is computationally indistinguishable from the view of $\adv$.
We assume the adversary $\adv$ that controls $\agg$, $\lambda_\user$ fraction of users and $\lambda_\node$ fraction of assisting nodes and its view consists of the joint view of the corrupted server $\agg^*$, the set of corrupted users $\calP_C $ and corrupted assisting nodes $\calA_C$: 
\begin{align*}
\mathsf{REAL}^{\Gamma (\calP, \calA, \agg, T)}_{\mathcal{S}^*, \calP_C, \calA_C, t}(\adv, \{\Vec{x}_i\}_{i\in\mathcal{P} \setminus \mathcal{C}_\user})
\approx_\kappa
\\
\mathsf{Simul}^{\Gamma (\calP, \calA, \agg, T), \mathcal{F}_{\mathcal{X}^t,\alpha}(\cdot)_{\{\Vec{x}_i\}_{i\in\mathcal{P} \setminus \calP_C}}}(\adv) 
\end{align*}
\end{definition}

\subsubsection{Integrity}\label{sec:threatmodelintegrity}

A malicious server can easily inject the final model with malicious updates.
In our threat model for model integrity, we consider a malicious and non-colluding server that aims to dishonestly aggregate user updates.
The non-colluding assumption is needed since colluding users can falsely claim their data as legitimate.\footnote{In the FL setting, clients can send random or malicious weights, resulting in a model susceptible to backdoor attacks~\cite{aistats/BagdasaryanVHES20}.
Mitigating this challenge with secure aggregation is difficult, as user gradients are masked throughout the training process, making it challenging to ensure computational correctness when the inputs appear random.
For example, assuming a trusted verifier or using heavyweight tools such as ZK-SNARK.
}

\section{Our Secure Aggregation Protocol}
Algorithms~\ref{alg:nanoSetup} and~\ref{alg:nanoAgg} present our \nano construction.
We consider the local updates $\Vec{w}_t^{\user_i}$ as high-dimensional vectors instead of individual coordinates as in many existing works (\eg,~\cite{popets/GuoPSBB24}).
The \mal{\uline{red and underlined parts are for malicious security}} and are not needed in the semi-honest setting.
Similarly, the \integrity{\dashuline{blue and dashed-underlined parts are for proof of honest aggregation.}}
\nano utilizes authenticated commitment (Definition~\ref{def:apvc}) and authenticated symmetric encryption (Definition~\ref{def:ae}) for integrity.


\begin{algorithm}[t]
\caption{Setup phase of \nano}\label{alg:nanoSetup}
\raggedright{
\textbf{Primitives:} 
A key exchange protocol $\Sigma$: $(\kakeygen, \kaagree)$.
\\
\mal{\uline{A digital signature scheme $\Pi$: $(\sgnkeygen, \sgnsign, \sgnverify)$.}}
\\
\integrity{\dashuline{An authenticated commitment scheme $\APVC$.}}
\\
\integrity{\dashuline{An authenticated symmetric encryption scheme $\mathtt{AE}$.}}
}

\raggedright{\textbf{Input:}
Security parameter $\kappa$;
Number of assisting nodes $k$.\\
All cryptographic primitives are instantiated using the security parameter $\kappa$.
\\
\integrity{\dashuline{Public parameters of $\APVC$:
$\pp = (\vec{g} = (g_1, \ldots, g_d), h)$.}}

\textbf{Output:}
All users receive the public keys of all assisting nodes $\langle\mal{\uline{(\pk^{\node_1}_\Pi, \dots, \pk^{\node_k}_\Pi),}}(\pk^{\node_1}_\Sigma, \dots, \pk^{\node_k}_\Sigma)\rangle$, and common secret seed $\x^{\user_i}_{\node_j}$ for each assisting node $\node_j$.
\\
All the assisting nodes receive the public keys of all participating users $\langle(\mal{\uline{\pk^{\user_1}_\Pi, \dots, \pk^{\user_n}_\Pi),}}(\pk^{\user_1}_\Sigma, \dots, \pk^{\user_n}_\Sigma)\rangle$, and common secret seed $\x^{\node_j}_{\user_i}$ with each user $\user_i$.
\\
\mal{ \uline{The aggregation server $\agg$ gets the public keys of all users and assisting nodes $\langle(\pk^{\user_1}_\Pi, \dots, \pk^{\user_n}_\Pi),(\pk^{\node_1}_\Pi, \dots, \pk^{\node_k}_\Pi)\rangle$.}}}\\

\hrulefill \vspace{-1mm}
 
\textbf{{Phase}} 1
\textbf{(Key Generation)}
\vspace{-2mm}

\hrulefill

All the communications below are conducted via an authenticated channel (similar to~\cite{ccs/BonawitzIKMMPRS17,popets/GuoPSBB24})

\begin{algorithmic}[1] 

\item Each user $\user_i$ generates its key pair(s) $(\sk^{\user_i}_\Sigma,\pk^{\user_i}_\Sigma) \gets \Sigma.\kakeygen(\kappa)$,
\mal{\uline{$(\sk^{\user_i}_\Pi,\pk^{\user_i}_\Pi) \gets \Pi.\sgnkeygen(\kappa)$}} 
and sends the keys $(\mal{\pk^{\user_i}_\Pi,}\pk^{\user_i}_\Sigma)$ to the $k$ assisting nodes \mal{\uline{and $\pk^{\user_i}_\Pi$ to $\agg$}}.
 
\item Each assisting node $\node_j$ generates its key pair(s) $(\sk^{\node_j}_\Sigma,\pk^{\node_j}_\Sigma) \gets \Sigma.\kakeygen(\kappa)$,
\mal{\uline{$(\sk^{\node_j}_\Pi,\pk^{\node_j}_\Pi) \gets \Pi.\sgnkeygen(\kappa)$}} 
and sends $(\mal{\pk^{\node_j}_\Pi,}\pk^{\node_j}_\Sigma)$ to the $n$ users \mal{ \uline{and sends $\pk^{\node_j}_\Pi$ the aggregation server $\agg$}}.

\item With $\pk^{\user_i}_\Sigma$ from $\user_i$, each assisting node $\node_j$ derives common secret seed $\x^{\node_j}_{\user_i} \gets \Sigma.\kaagree(\sk^{\node_j}_{\Sigma},\pk^{\user_i}_\Sigma) $.
\item With $\pk^{\node_j}_\Sigma$ from assisting node $\node_j$, the user $\user_i$ derives common secret seed  $\x^{\user_i}_{\node_j} \gets \Sigma.\kaagree(\sk^{\user_i}_{\Sigma},\pk^{\node_j}_{\Sigma}) $.

\item \integrity{\dashuline{One assisting node executes $\rho \gets \APVC.\KeyGen(1^\secParam) $, computes $c_{\node_j}^{\user_i} \gets \aeenc(\x_{\node_j}^{\user_i},\rho)$ for all $i\in\{1,\dots,n\}$, and distribute $c_{\node_j}^{\user_i}$ to the $n$ users.}}
\item \integrity{\dashuline{With $c_{\node_j}^{\user_i}$, each user $\user_i$ recovers $\rho \gets \aedec(\x_{\user_i}^{\node_j},c_{\node_j}^{\user_i})$.}}

\item \mal{\uline{Aggregation server generates a key pair $(\sk^\mathtt{S}_\Pi,\pk^\mathtt{S}_\Pi)$
\\
$\gets \Pi.\sgnkeygen(1^\kappa)$ and sends $\pk^\mathtt{S}_\Pi$ to all users.}}

\end{algorithmic}

\end{algorithm}
\begin{algorithm}[!t]
\caption{Aggregation phase of \nano}\label{alg:nanoAgg}
\textbf{Primitives:} 
A pseudorandom function family $\PRF(\cdot)$.
\\
\mal{\uline{A digital signature scheme $\Pi$: $(\sgnkeygen, \sgnsign, \sgnverify)$.}}
\\
\integrity{\dashuline{An authenticated commitment scheme $\APVC$.}}
\\
\integrity{\dashuline{An authenticated symmetric encryption scheme $\mathtt{AE}$.}}

\raggedright{\textbf{Input:}
Iteration number $t$.
User-derived model update~$\vec{w}^{\user_i}_t$.
\\
Common secret seeds $\x$ (of each user-assisting-node pair).
\\
\mal{\uline{The signature key pair}}.
 
\textbf{Output:} The final model update $\vec{w}_t \in \mathcal{M}^d$.
\\

\hrulefill
\\
\textbf{{Phase}} 1 \textbf{(Masking Updates)}}
\\\vspace{-2pt}
\hrulefill

\begin{algorithmic}[1]
\State 
User $\user_i$ computes masked update
$\vec{y}^{\user_i}_{t}
= \vec{w}^{\user_i}_{t} + \Vec{a}^{\user_i}_{t}$,
where $\Vec{a}^{\user_i}_{t} = \sum_{j = 1}^k \PRF(\x^{\user_i}_{\node_j}, t)$

\State
\integrity{\dashuline{User $\user_i$ computes $\cm^{\user_i}_t \gets \APVC.\Comm(\rho, \vec{w}^{\user_i}_{t}, \vec{\Vec{a}}^{\user_i}_{t}[0])$}}.

\State
User sets $m = (t, \vec{y}^{\user_i}_{t}, \integrity{\dashuline{\cm^{\user_i}_t}})$ and $m' = (t)$,
\mal{\uline{signs them via 
$\sigma^{\user_i}_{\agg} \gets \Pi.\sgnsign(\sk^{\user_i}_\Pi,m)$, 
$\sigma^{\user_i}_{\node_j} \gets \Pi.\sgnsign(\sk^{\user_i}_\Pi,m')$}} and sends 
$(m\mal{\uline{, \sigma^{\user_i}_{\agg}}})$ and 
$(m'\mal{\uline{, \sigma^{\user_i}_{\node_j}}})$ 
to the aggregation server and assisting nodes $\node_j, \forall j \in [k]$, respectively.
\end{algorithmic}

\hrulefill
\\
\textbf{{Phase}} 2 \textbf{(Aggregate Updates)}
\\\vspace{-2pt}
\hrulefill
 
\begin{algorithmic}[1]
\State
Upon receiving $(m'\mal{\uline{, \sigma^{\user_i}_{\node_j}}})$
from all the system users,
$\node_j$ \mal{\uline{checks if $\Pi.\sgnverify(\pk_\Pi^{\user_i},m', \sigma^{\user_i}_{\node_j} ) \stackrel{\smash{?}}{=} 1$ holds; if so, it}} adds the user to its user list $\mathcal{L}_{j,t}$.

\State
Each assisting node checks if {$|\mathcal{L}_{j,t}| \geq \alpha \calP_H$};
if so, it computes $\Vec{a}^{\node_j}_{t}=\sum_{i=1}^{|\mathcal{L}_{j,t}|}\PRF(\x^{\node_j}_{\user_i},t)$, sets $m^{\prime\prime} =(t,|\mathcal{L}_{j,t}|, \Vec{a}^{\node_j}_t)$ \mal{\uline{and $\sigma^{\node_i}_{\agg} \gets \Pi.\sgnsign(\sk^{\node_j}_\Pi,m'')$ }} and sends $(m''\mal{\uline{, \sigma^{\node_j}_{\agg}}})$ to the aggregation server $\agg$.

\State
With $(m\mal{\uline{, \sigma^{\user_i}_{\agg}}})$, 
$\agg$ \mal{\uline{checks if 
$\Pi.\sgnverify(\pk_\Gamma^{\user_i}, m, \sigma^{\user_i}_{\agg} )\stackrel{\smash{?}}{=} 1$\!}}
\mal{\uline{and if so, it}}
adds $\user_i$ to its user list $\mathcal{L}_{\agg, t}$.

\State
\mal{\uline{$\agg$ checks if
$\Pi.\sgnverify(\pk_\Pi^{\node_j},m'', \sigma^{\node_j}_{\agg})
\stackrel{{?}}{=} 1$
$\forall j \in [k]$.
}}
It~continues if $\mathcal{L}_{\agg, t} = \mathcal{L}_{1, t} = \dots = \mathcal{L}_{k, t}$
and then broadcasts update $\vec{w}_t = \sum_{i = 1}^{|\mathcal{L}_{\agg,t}|}\vec{y}^{\user_i}_{t} - \sum_{j =1 }^k \Vec{a}^{\node_j}_t$.

\State
\integrity{\dashuline{For all the users in
$|\mathcal{L}_{\agg,t}|$, 
$\agg$ computes sends $(x_t, \vec{w}_t)$ to online users,
where $x_t = \prod_{i=1}^{n} \cm^{\user_i}_t \prod_{j=1}^{k}h^{-\vec{\Vec{a}}_t^{A_j}[0]}$.
}} 

\State
\integrity{\dashuline{All user $\user_i$ accepts $\vec{w}_t$ if $\vec{g}^{\vec{w}_t\cdot \rho} = x_t$.}}

\end{algorithmic}

\end{algorithm}

\subsection{Detailed Construction}
Our approach extends the assumption made in most existing protocols, which rely on a subset of honest users, to a set of assisting nodes.
For privacy guarantee, only one assisting needs to be honest.
As discussed later (Section~\ref{sec:experiments}), the computation and communication overhead for assisting nodes are on par with common users in some of the most efficient existing protocols.
Thus, we can assume a dynamic and rotating set of users (similar to~\cite{ccs/DavidM0NT22}) to serve as assisting nodes in each iteration (see Section~\ref{sec:dynamic}).

In Algorithm~\ref{alg:nanoSetup},
each user generates a $(\sk^{\user_i}_\Sigma, \pk^{\user_i}_\Sigma)$ pair for the key exchange protocol to compute a common secret $\x_{\user_i}^{\node_j} $ with assisting nodes $\node_j$.
In the malicious setting,
users and assisting nodes set up key pairs for signatures.
$\APVC$ is also initialized,
and a system secret $\rho$ is computed, encrypted under each user secret, and sent to all users.

In the first round of Algorithm~\ref{alg:nanoAgg},
updates are masked for aggregation.
Each user computes a masking vector $\Vec{a}_t^{\user_i}$
(using common secrets $\x_{\user_i}^{\node_j} $ for $j = \{1, \dots, k\}$) to mask their local gradient vector ${\Vec{w}_t^{\user_i}}$.
For model integrity, each user uses $\APVC.\Comm(\cdot)$ to derive a commitment $\cm^{\user_i}$ on their local gradient and sends it to aggregation server $\agg$.
The masked gradient $\Vec{y}_t^{\user_i}$ and 
the iteration number
are then sent to $\agg$ and the $k$ assisting nodes, respectively.
In the malicious setting, all outgoing messages are signed
using $\Pi.\sgnsign(\cdot)$.

The second round is for aggregate updates.
In the semi-honest (malicious) setting, upon receiving (and verifying) the participation message (and signature), each $\node_j$ adds the user to the list $\mathcal{L}_{j, t}$.
Then for all users in $\mathcal{L}_{j,t}$, it computes the aggregation of all their masking terms $\Vec{a}^{\node_j}_t$ (using the common secret keys $\x^{\node_j}_{\user_i}$) and sends 
$
\vec{y}^{\node_j}_{\user_i} = \Vec{a}^{\node_j}_{t} + \x^{\node_j}_{\user_i}$
to the aggregation server.
In the malicious setting, again, assisting nodes sign all their outgoing messages using $\Pi.\sgnsign(\cdot)$.

Next, in the semi-honest (malicious) setting, upon receiving (and verifying) the participating message (and signature), $\agg$ adds the user to the list $\mathcal{L}_{\agg, t}$.
It then checks if all the user lists (of assisting nodes and the aggregation server) match.
Then, for all the users in $\mathcal{L}_{\agg,t}$, the server uses the masked updates ($\Vec{y}_t^{\user_i}$) and the aggregated masking terms ($\Vec{a}^{\node_j}_t$), supplied by the assisting nodes, to compute the aggregation.
For model integrity, $\agg$ computes~$x$ using the users' commitments $\cm^{\user_i}_t$.
Lastly, $\agg$ returns the aggregation result (and $x$ for the proof of integrity)
back to the users.

\subsubsection{Extra Mechanisms around User Dropout}
To ensure consistency, a dispute phase can be introduced.
If a user is missing from the list of a certain assisting node, the missing user participation message (and signature) can be delivered to the assisting node to update their list.
Alternatively, We could assume a star network topology, where users send everything through the aggregation server, which then distributes them to the assisting nodes.

\subsubsection{Dynamic Selection of Assisting nodes}\label{sec:dynamic}
\nano can be instantiated with as few as two assisting nodes (\ie, $k = 2$)
while maintaining privacy as long as at least one node is honest.
Other protocols relying on external assistance to unmask the gradients require a much higher participation rate from these nodes
(\eg, Flamingo~\cite{sp/MaWAPR23} expects ${\geq}64$ decryptors per iteration).
\nano presented in Algorithms \ref{alg:nanoSetup} and \ref{alg:nanoAgg} assumes a static set of assisting nodes.
However, by incorporating randomness from a trusted source (\eg,~\cite{sp/DasKI022}), the protocol can initialize with $y \gg k$ assisting nodes during the setup phase and then randomly sample $k$ of them for each iteration.
Assisting nodes need not be high-performance devices.
For instance, during our aggregation phase with $200$ users, assisting nodes require only ${\sim}1.5$ GB of memory, even when handling large models with $16,000$ weights.

\subsubsection{Resiliency Against Assisting Nodes Dropout}\label{sec:Resiliency}

The failure of an assisting node can impede model convergence, resulting in wasted training iterations.
We suggest two approaches to enhance resiliency.
First, by dynamically selecting a small set of assisting nodes in each iteration and employing a simple secret-sharing method, other assisting nodes can reconstruct the dropped node's secret and assist the server in unmasking the model.
This would only incur an overhead in terms of $k$, which is usually a small number and does not introduce any overhead on the users.
However, it might expose the secret of the dropped assisting node, disqualifying it from future selection.

Alternatively, we can use key-homomorphic PRF\mbox{~\cite{crypto/BonehLMR13}}.
This method permits the computation of the dropped node's unmasking term $\mathbf{a}_{t}^{\node_j}$ using only secret shares, thereby preserving the confidentiality of the assisting node's secret key.


\subsection{Security Analysis}
The proofs for all lemmas and theorems in this section are deferred to the appendix.

\begin{theorem}\label{thm:correctness}
Given a secure key exchange protocol (Definition~\ref{def:keyexchange}), \nano protocol presented in Algorithms~\ref{alg:nanoSetup} and~\ref{alg:nanoAgg}, with parameters $(1-\delta)\geq\alpha$ guarantees correctness with $\delta$ offline rate.
\end{theorem}

\begin{lemma}
Our \nano protocol is resistant to user dropout up as long as $\alpha|\mathcal{P}_h|$ users participate.

\end{lemma}

\myPara{Privacy Against Semi-Honest Adversary}\label{sec:securitysemi-honest}
Following~\cite{ccs/BonawitzIKMMPRS17,ccs/BellBGL020,popets/GuoPSBB24}, we consider two scenarios.
In the first one, the adversary only controls a subset of the users and assisting nodes, but not the server.
The proof of such a setting is rather trivial since the joint view of any subset of the users and assisting nodes is fully independent of other entities in the system.
The second scenario considers the adversary that controls the aggregation server, as well as a subset of users and assisting nodes.
We prove the security of our protocol against such a semi-honest adversary in the following.
\begin{theorem}\label{thm:semi-honest}
	The \nano protocol presented in Algorithms~\ref{alg:nanoSetup} and~\ref{alg:nanoAgg}, running with $n$ parties
	$\{\user_1, \ldots, \user_n\}$, $k$ assisting nodes
	$\{\node_1, \ldots, \node_k\}$, and an aggregation server $\agg$ provides privacy against a semi-honest adversary $\adv$ which controls $\agg$ and $1-\alpha$ fraction of users and $k - 1$ assisting nodes with the offline rate $(1 - \delta) \geq \alpha$.
\end{theorem}

\myPara{Privacy Against Malicious Adversaries}
Similar to the above, we consider two possible scenarios for the malicious setting.
In the first scenario, the malicious adversary $\adv$ does not have control over the server $\agg$.
In this case, privacy can be easily proven using the same rationale as above.
Therefore, in the following, we focus on an $\adv$ that controls $\agg$ along with a subset of the users and assisting nodes.

\begin{theorem}\label{thm:malicious}
	The \nano protocol presented in Algorithms~\ref{alg:nanoSetup} and~\ref{alg:nanoAgg}, running with $n$ parties 
	$\{\user_1, \ldots, \user_n\}$, $k$ assisting nodes 
	$\{\node_1, \ldots, \node_k\}$, and an aggregation server $\agg$ provides privacy against a malicious adversary $\adv$ which controls $\agg$ and $1 - \alpha$ fraction of users and $k - 1$ assisting nodes with the offline rate $(1 - \delta) \geq \alpha$.
\end{theorem}

\begin{lemma}\label{lemma:integrity}
The protocol in Algorithms~\ref{alg:nanoSetup} and~\ref{alg:nanoAgg} with $\APVC$ offers local model's privacy and aggregation integrity against semi-honest or malicious adversary.
\end{lemma}

\begin{table*}[!t]
	\centering
	 \caption{Analytical Performance of Setup Phase}
	\label{tab:analytical_setup} 
	\begin{threeparttable}
	\begin{tabular}{|c|c|c|c|}
			\hline
		Protocol
		 & \textbf{User} &
		
		 \textbf{Assisting Node/Decryptor} & \textbf{Server} 
		\\ 	\hline
		  \cite{ccs/BellBGL020} & N/A & N/A & N/A \\\hline

\cite{popets/GuoPSBB24} &
		\begin{tabular}{@{}c@{}c@{}} $1_\Kg + 1_\Sgn+|U'-1|(1_\Vfy+$
		\\
		$1_\Ka + 1_{\sshl} + 1_\AEnc+1_\ADec)$
		\end{tabular} & N/A
		& ${|U|_\Vfy }$
		\\	\hline

		\cite{sp/MaWAPR23} &
		$|D'|_\Vfy$
		& \begin{tabular}{@{}c@{}c@{}}$2_{\PIL} + 3_\Sgn+2_{\sshl} +$	\\ $	|D|(3|D'|+1)_\Vfy + $\\
		$(|D'|\cdot|D|)_\Mu$\end{tabular}
		& $(|D'|\cdot|D|)_\Vfy$ 
		\\	\hline

		 Ours &
		 $2_\Kg + k_\Ka$
		 & $2_\Kg + k_\Ka$
		 & $1_\Kg$\\
		 \hline
		 		 
	\end{tabular}
\begin{tablenotes}[para, flushleft]

	$\PIL$ denotes polynomial interpolation of length $L$.
	$\sshl$ denote secret sharing operation to $l$ shares.
	$\Mu$ denotes multiplication, and $\AEnc$, $\ADec$,  denote asymmetric encryption and decryption.

	\end{tablenotes}
	\end{threeparttable}
\end{table*}

\begin{table*}[!t]
	\centering
	 \caption{Analytical Performance of Aggregate Phase}
	\label{tab:analytical_agg} 
	\begin{threeparttable}
	\begin{tabular}{|c|c|c|c|}
			\hline
		
Protocol		 & \textbf{User} &
		 \textbf{Assisting Node/Decryptor}& \textbf{Server} \\	\hline
		  \cite{ccs/BellBGL020} &\begin{tabular}{@{}c@{}c@{}}
$2_\Kg+2_{\sshl} +\log n (1_\SEnc+1_\SDec + $ \\
${1_\Ka + 2_\Sgn+2_\Vfy})+(2\log n)_\Su$ \end{tabular}& N/A &$1_\Rssh + 2\log n_\Su$\\\hline

\cite{popets/GuoPSBB24} 
		&\begin{tabular}{@{}c@{}c@{}} $(|\vec{w}|+1)_\Ex+1_\Sgn+$\\$|U-1|(1_\Su+1_\Vfy)+1_\Su$ \end{tabular}& N/A & \begin{tabular}{@{}c@{}c@{}} $1_{\Rsshe} +|\vec{w}|_\Ex + $\\
		$1_\Mu$\end{tabular}	\\	\hline

		\cite{sp/MaWAPR23} & \begin{tabular}{@{}c@{}c@{}} $2_\PRG+|D'|(1_\Ex+1_\Hash+1_\PRF+$\\$1_\Su+1_\AEnc+1_\Sgn+1_\SEnc)+1_{\sshl}$ \end{tabular}&\begin{tabular}{@{}c@{}c@{}}$1_\Sgn+|U'|_\SDec +$ \\ $(|D|+|U'|)_\Vfy
		$\end{tabular}
		&
		\begin{tabular}{@{}c@{}c@{}} $|D|_{\PIL}+|U'|(1_\Su + $ \\ $1_\Mu +1_\Rssh)+2_\PRG+ $ \\ $ (|U'|+1)_\Su $ \end{tabular}
		\\	\hline

		 Ours 
		 & $k_\PRF+|k+1|_\Su+2_\Sgn$ &\begin{tabular}{@{}c@{}c@{}} $|U'| (1_\Vfy+1_\Su+ $\\
		 $1_\PRF)+ {1_\Sgn}$\end{tabular}
		 & \begin{tabular}{@{}c@{}c@{}}$(|U'|+|k|)(1_\Vfy + $ \\ $1_\Su)$
		 \end{tabular}\\
		 \hline
		 		 
	\end{tabular}
\begin{tablenotes}[para, flushleft]

	$\Rssh$, and $\Rsshe$ denote share reconstruction and share reconstruction in the exponent.
$\Su$ and $\Ex$ denote summation and exponentiation operations.
	$\SEnc$, and $\SDec$ denote symmetric encryption and decryption.

	\end{tablenotes}
	\end{threeparttable}
\end{table*}


\begin{figure*}[!t]
	\centering
		\resizebox{1\textwidth}{!}{
			\begin{subfigure}{.3\textwidth}
%
%
\definecolor{A}{HTML}{e6194B}%
\definecolor{B}{HTML}{f58231}%
\definecolor{C}{HTML}{4363d8}%
\definecolor{D}{HTML}{911eb4}%
\definecolor{E}{HTML}{3cb44b}%
\definecolor{F}{rgb}{0.92900,0.69400,0.12500}%
\definecolor{G}{HTML}{808000}%
\definecolor{H}{HTML}{000000}%
\begin{tikzpicture}
	\begin{axis}[%
		width=.8\textwidth,
		height=0.5\textwidth,
		at={(1.128in,0.894in)},
		scale only axis,
		xmin=0,
		xmax=6,
		xlabel={Number of users},
		xtick={0, 1, 2, 3, 4, 5, 6},
		xticklabels={$ $, $200$, $ 400 $, $ 600 $, $ 800 $, $ 1000 $},
            xlabel style={font=\footnotesize},
            ylabel style={font=\footnotesize},
            tick label style={font=\footnotesize},
		ymax=10000,
            ymode=log,
		ylabel = {Delay (ms) (log)},
		ylabel shift=-8pt,
		yticklabel shift={0cm},
		axis background/.style={fill=white},
		legend columns=2,
		legend style={legend cell align=left, align=left, fill=none, draw=none,inner sep=-0pt, row sep=0pt, font = \tiny},
		legend pos = north west,
		%
		minor y tick num=5,
		]

		\addplot [color=B, dashed, mark=square*, mark options={solid, B}]
		table[row sep=crcr]{%
			1 10.1010\\ 
			2 46.4646\\ 
			3 123.2323\\ 
			4 242.4242\\ 
			5 375.7575\\ 
		};
		\addlegendentry{MicroSecAgg$_1$}

		\addplot [color=C, loosely dotted, mark=triangle*, mark options={solid, C}]
		table[row sep=crcr]{%
			1 22.2222\\ 
			2 52.5252\\ 
			3 90.9090\\ 
			4 133.3333\\ 
			5 177.7777\\ 
		};
		\addlegendentry{MicroSecAgg$_2$(50)}

		\addplot [color=H, dashed, mark=pentagon*, mark options={solid, H}]
		table[row sep=crcr]{%
			1 0.15091896057128906\\ %
			2 0.15091896057128906\\ %
			3 0.15091896057128906\\ %
			4 0.15091896057128906\\ %
			5 0.15091896057128906\\ 
		};
		\addlegendentry{\FlamingoServerFirst}

		\addplot [color=D, loosely dotted, mark=otimes*, mark options={solid, D}]
		table[row sep=crcr]{%
			1 0.3190040588378906\\ %
			2 0.3190040588378906\\ %
			3 0.3190040588378906\\ %
			4 0.3190040588378906\\ %
			5 0.3190040588378906\\ 
		};
		\addlegendentry{\FlamingoServerSec}

		\addplot [color=A, solid, mark=diamond*, mark options={solid, A}]
		table[row sep=crcr]{%
			1 0.1900\\ 
			2 0.1900\\
			3 0.1900\\ 
			4 0.1900\\ 
			5 0.1900\\
		};
		\addlegendentry{\nano~(MS)}

    \draw[solid, black, latex-latex, line width=0.01pt] (515, -1) -- (515,5.984907657724967);
    \node[solid, black, latex-latex] at (549, 2.1568306293) {\tiny 1178x};

    \draw[solid, black, latex-latex, line width=0.01pt] (515, -1.691246399047170) -- (515,-1);
    \node[solid, black, latex-latex] at (545, -1.3380770284) {\tiny 1.7x};
    
	\end{axis}

\end{tikzpicture}%
					\caption{Server}\label{fig:exp:IMS:server} 
			\end{subfigure}
			\begin{subfigure}{.3\textwidth}
%
%
\definecolor{A}{HTML}{e6194B}%
\definecolor{B}{HTML}{f58231}%
\definecolor{C}{HTML}{4363d8}%
\definecolor{D}{HTML}{911eb4}%
\definecolor{E}{HTML}{3cb44b}%
\definecolor{F}{rgb}{0.92900,0.69400,0.12500}%
\definecolor{G}{HTML}{808000}%
\definecolor{H}{HTML}{000000}%
\begin{tikzpicture}
	\footnotesize
	\begin{axis}[%
		width=.8\textwidth,
		height=0.5\textwidth,
		at={(1.128in,0.894in)},
		scale only axis,
		xmin=0,
		xmax=6,
		xlabel={Number of users},
		xtick={0, 1, 2, 3, 4, 5, 6},
		xticklabels={$ $, $200$, $ 400 $, $ 600 $, $ 800 $, $ 1000 $},
            xlabel style={font=\footnotesize},
            ylabel style={font=\footnotesize},
            tick label style={font=\footnotesize},
		ymax=1000000000,
            ymode=log,
		ylabel = {Delay (ms) (log)},
		ylabel shift=-5pt,
		yticklabel shift={0cm},
		axis background/.style={fill=white},
		legend columns=2,
		legend style={legend cell align=left, align=left, fill=none, draw=none,inner sep=-0pt, row sep=0pt, font = \tiny},
		legend pos = north west,
		%
		minor y tick num=5,
		]
          
        \addplot [color=E, solid, mark=asterisk, mark options={solid, E}]
		table[row sep=crcr]{%
                1 0.1889 \\ 
			2 0.1889 \\ 
			3 0.1889\\ 
			4 0.1889 \\ 
			5 0.1889 \\ 
		};
		\addlegendentry{\nano~(SH)}
  
		\addplot [color=A, solid, mark=diamond*, mark options={solid, A}]
		table[row sep=crcr]{%
                1 0.2519 \\ 
			2 0.2519\\ 
			3 0.2519\\
			4 0.2519\\
			5 0.2519\\
		};
		\addlegendentry{\nano~(MS)}
		\addplot [color=B, dashed, mark=square*, mark options={solid, B}]
		table[row sep=crcr]{%
			1 88.785\\ 
			2 350.467\\ 
			3 855.140\\ 
			4 1584.112\\ 
			5 2509.345\\ 
		};
		\addlegendentry{MicroSecAgg$_1$}

		\addplot [color=H, dashed, mark=pentagon*, mark options={solid, H}]
		table[row sep=crcr]{%
			1 26.7369747162\\ %
			2 26.7369747162\\ %
			3 26.7369747162\\ %
			4 26.7369747162\\ %
			5 26.7369747162\\ 
		};
		\addlegendentry{\FlamingoUserFirst}

		\addplot [color=D, loosely dotted, mark=otimes*, mark options={solid, D}]
		table[row sep=crcr]{%
			1 53.49111557\\ %
			2 53.49111557\\ %
			3 53.49111557\\ %
			4 53.49111557\\ %
			5 53.49111557\\ 
		};
		\addlegendentry{\FlamingoUserSec}

            \addplot [color=C, loosely dotted, mark=triangle*, mark options={solid, C}]
		table[row sep=crcr]{%
			1 32.710\\ 
			2 32.710\\ 
			3 32.710\\ 
			4 32.710\\ 
			5 32.710\\ 
		};
		\addlegendentry{MicroSecAgg$_2$(50)}

		\addplot [color=teal, dashed, mark=halfcircle*, mark options={solid, teal}]
		table[row sep=crcr]{%
			1 600.0770195007\\ 
			2 600.0770195007\\ 
			3 600.0770195007\\ 
			4 600.0770195007\\ 
			5 600.0770195007\\ 
		};
		\addlegendentry{\FlamingoDecFirst}

		\addplot [color=magenta, loosely dotted, mark=halfsquare*, mark options={solid, magenta}]
		table[row sep=crcr]{%
			1 2416.2848447418\\ 
			2 2416.2848447418\\ 
	    	3 2416.2848447418\\ 
			4 2416.2848447418\\ 
			5 2416.2848447418\\ 
		};
		\addlegendentry{\FlamingoDecSec}


        \draw[solid, black, latex-latex, line width=0.01pt] (515, 3.384907657724967) -- (515,7.784907657724967);
    \node[solid, black, latex-latex] at (544, 5.1568306293) {\tiny 94x};

        \draw[solid, black, latex-latex, line width=0.01pt] (515, -1.652463990471709) -- (515,3.384907657724967);
    \node[solid, black, latex-latex] at (545, 1.008306293) {\tiny 142x};
  
        \end{axis}

\end{tikzpicture}%
					\caption{User}\label{fig:exp:IMS:user} 
			\end{subfigure}
			\begin{subfigure}{.3\textwidth}
%
%
\definecolor{A}{HTML}{e6194B}%
\definecolor{B}{HTML}{f58231}%
\definecolor{C}{HTML}{4363d8}%
\definecolor{D}{HTML}{911eb4}%
\definecolor{E}{HTML}{3cb44b}%
\definecolor{F}{rgb}{0.92900,0.69400,0.12500}%
\definecolor{G}{HTML}{808000}%
\definecolor{H}{HTML}{000000}%
\begin{tikzpicture}
	\footnotesize
	\begin{axis}[%
		width=.8\textwidth,
		height=0.5\textwidth,
		at={(1.128in,0.894in)},
		scale only axis,
		xmin=0,
		xmax=6,
		xlabel={Number of users},
		xtick={0, 1, 2, 3, 4, 5, 6},
		xticklabels={$ $, $200$, $ 400 $, $ 600 $, $ 800 $, $ 1000 $},
		ymin= 5,
		ymax=60,
		ylabel = {Delay (ms)},
		ylabel shift=-5pt,
		yticklabel shift={0cm},
		axis background/.style={fill=white},
		legend columns=2,
		legend style={legend cell align=left, align=left, fill=none, draw=none,inner sep=-0pt, row sep=0pt, font = \tiny},
		legend pos = north west,
		%
		minor y tick num=5,
		]

       \addplot [color=A, solid, mark=diamond*, mark options={solid, A}]
		table[row sep=crcr]{%
                1 11.0790\\ 
			2 21.0562\\  
			3 31.7832\\ 
			4 41.9725\\ 
			5 51.7883\\
		};
		\addlegendentry{\nano~(MS)}

        \addplot [color=E, solid, mark=asterisk, mark options={solid, E}]
		table[row sep=crcr]{%
                1 11.0101 \\ 
			2 21.0175\\  
			3 31.1398\\ 
			4 41.2646\\ 
			5 50.8841\\ 
		};
		\addlegendentry{\nano~(SH)}

	\end{axis}

\end{tikzpicture}%
					\caption{Assisting node} \label{fig:exp:IMS:node}
			\end{subfigure}
		} 

	\caption{Computation in the Setup phase}\label{fig:exp:IMS} 
	\vspace{-10pt}
\end{figure*}

\begin{figure*}[!t]
	\centering
		\resizebox{1\textwidth}{!}{
			\begin{subfigure}{.3\textwidth}
%
%
\definecolor{A}{HTML}{e6194B}%
\definecolor{B}{HTML}{f58231}%
\definecolor{C}{HTML}{4363d8}%
\definecolor{D}{HTML}{911eb4}%
\definecolor{E}{HTML}{3cb44b}%
\definecolor{F}{rgb}{0.92900,0.69400,0.12500}%
\definecolor{G}{HTML}{808000}%
\definecolor{H}{HTML}{000000}%
\begin{tikzpicture}
	\footnotesize
	\begin{axis}[%
		width=.8\textwidth,
		height=0.5\textwidth,
		at={(1.128in,0.894in)},
		scale only axis,
		xmin=0,
		xmax=6,
		xlabel={Number of users},
		xtick={0, 1, 2, 3, 4, 5, 6},
		xticklabels={$ $, $200$, $ 400 $, $ 600 $, $ 800 $, $ 1000 $},
            xlabel style={font=\footnotesize},
            ylabel style={font=\footnotesize},
            tick label style={font=\footnotesize},
		ymax=1000000000000,
            ymode=log,
		ylabel = {Bandwidth (B) (log)},
		ylabel shift=-5pt,
		yticklabel shift={0cm},
		axis background/.style={fill=white},
		legend columns=2,
		legend style={legend cell align=left, align=left, fill=none, draw=none,inner sep=-0pt, row sep=0pt, font = \tiny},
		legend pos = north west,
		%
		minor y tick num=5,
		]
		\addplot [color=A, solid, mark=diamond*, mark options={solid, A}]
		table[row sep=crcr]{%
            1 6600 \\ 
			2 13200 \\ 
			3 19800\\ 
			4 26400\\ 
			5 33000\\ 
		};

		\addlegendentry{\nano~(MS)}

  		\addplot [color=B, dashed, mark=square*, mark options={solid, B}]
		table[row sep=crcr]{%
			1 40000000\\ 
			2 171428571.428\\ 
			3 377142857.142\\ 
			4 687619047.619\\ 
			5 1057142857.142\\ 
		};
		\addlegendentry{MicroSecAgg$_1$}

		 \addplot [color=C, loosely dotted, mark=triangle*, mark options={solid, C}]
		 table[row sep=crcr]{%
		 	1 49523809.523\\ 
		 	2 100952380.952\\ 
		 	3 150476190.476\\ 
		 	4 201904761.904\\ 
		 	5 251428571.428\\ 
		 };
		 \addlegendentry{MicroSecAgg$_2$(50)}

		\addplot [color=H, dashed, mark=pentagon*, mark options={solid, H}]
		table[row sep=crcr]{%
			1 9247488\\ 
			2 10898688\\ %
			3 12549888\\ %
			4 14201088\\ %
			5 15852288\\ %
		};
		\addlegendentry{\FlamingoServerFirst}

		\addplot [color=D, loosely dotted, mark=otimes*, mark options={solid, D}]
		table[row sep=crcr]{%
			1 38090240\\ %
			2 41392640\\ %
			3 44695040\\ %
			4 47997440\\ %
			5 51299840\\ %
		};
		\addlegendentry{\FlamingoServerSec}

    \draw[solid, black, latex-latex, line width=0.01pt] (515, 10.2912463990471709) -- (515,16.584907657724967);
    \node[solid, black, latex-latex] at (550, 13.5380770284) {\tiny 480x};

    \draw[solid, black, latex-latex, line width=0.01pt] (515, 16.584907657724967) -- (515,20.784907657724967);
    \node[solid, black, latex-latex] at (547, 18.5380770284) {\tiny 67x};

	\end{axis}

\end{tikzpicture}%
					\caption{Server}\label{fig:exp:BWS:server}
			\end{subfigure}
			\begin{subfigure}{.3\textwidth}
%
%
\definecolor{A}{HTML}{e6194B}%
\definecolor{B}{HTML}{f58231}%
\definecolor{C}{HTML}{4363d8}%
\definecolor{D}{HTML}{911eb4}%
\definecolor{E}{HTML}{3cb44b}%
\definecolor{F}{rgb}{0.92900,0.69400,0.12500}%
\definecolor{G}{HTML}{808000}%
\definecolor{H}{HTML}{000000}%
\begin{tikzpicture}
	\footnotesize
	\begin{axis}[%
		width=.8\textwidth,
		height=0.5\textwidth,
		at={(1.128in,0.894in)},
		scale only axis,
		xmin=0,
		xmax=6,
		xlabel={Number of users},
		xtick={0, 1, 2, 3, 4, 5, 6},
		xticklabels={$ $, $200$, $ 400 $, $ 600 $, $ 800 $, $ 1000 $},
            xlabel style={font=\footnotesize},
            ylabel style={font=\footnotesize},
            tick label style={font=\footnotesize},
		ymax= 1000000000,
        ymode=log,
		ylabel = {Bandwidth (B) (log)},
		ylabel shift=-5pt,
		yticklabel shift={0cm},
		axis background/.style={fill=white},
		legend columns=2,
		legend style={legend cell align=left, align=left, fill=none, draw=none,inner sep=-0pt, row sep=0pt, font = \tiny},
		legend pos = north west,
		%
		minor y tick num=5,
		]
  
        \addplot [color=E, solid, mark=asterisk, mark options={solid, E}]
		table[row sep=crcr]{%
            1 99 \\ 
			2 99 \\ 
			3 99 \\ 
			4 99 \\
			5 99 \\
		};
		\addlegendentry{\nano~(SH)}
  
		\addplot [color=A, solid, mark=diamond*, mark options={solid, A}]
		table[row sep=crcr]{%
                1 231\\ 
			2 231\\ 
			3 231\\
			4 231\\
			5 231\\
		};
		\addlegendentry{\nano~(MS)}

  \addplot [color=B, dashed, mark=square*, mark options={solid, B}]
		table[row sep=crcr]{%
			1 185826.7\\
			2 374803.1 \\
			3 552755.9\\
			4 749606.2\\
			5 927559.0\\
		};
		\addlegendentry{MicroSecAgg$_1$}

  \addplot [color=C, loosely dotted, mark=triangle*, mark options={solid, C}]
		table[row sep=crcr]{%
			1 187401.5\\
			2 187401.5\\
			3 187401.5\\
			4 187401.5\\
			5 187401.5\\
		};

		\addlegendentry{MicroSecAgg$_2$(50)}

		\addplot [color=teal, dashed, mark=halfcircle*, mark options={solid, teal}]
		table[row sep=crcr]{%
			1 2013\\ 
			2 2013\\ 
			3 2013\\ 
			4 2013\\ 
			5 2013\\ 
		};
		\addlegendentry{\FlamingoDecFirst}

		\addplot [color=magenta, loosely dotted, mark=halfsquare*, mark options={solid, magenta}]
		table[row sep=crcr]{%
			1 3612\\ 
			2 3612\\ 
			3 3612\\ 
			4 3612\\ 
			5 3612\\ 
		};
		\addlegendentry{\FlamingoDecSec}

    \draw[solid, black, latex-latex, line width=0.01pt] (515, 4.612463990471709) -- (515,7.734907657724967);
    \node[solid, black, latex-latex] at (545, 6.5236858241) {\tiny 20x};

        \draw[solid, black, latex-latex, line width=0.01pt] (515, 7.734907657724967) -- (515,13.784907657724967);
    \node[solid, black, latex-latex] at (553, 10.5236858241) {\tiny 461x};

	\end{axis}

\end{tikzpicture}%
					\caption{User}\label{fig:exp:BWS:user}
			\end{subfigure}
			\begin{subfigure}{.3\textwidth}
%
%
\definecolor{A}{HTML}{e6194B}%
\definecolor{B}{HTML}{f58231}%
\definecolor{C}{HTML}{4363d8}%
\definecolor{D}{HTML}{911eb4}%
\definecolor{E}{HTML}{3cb44b}%
\definecolor{F}{rgb}{0.92900,0.69400,0.12500}%
\definecolor{G}{HTML}{808000}%
\definecolor{H}{HTML}{000000}%
\begin{tikzpicture}
	\footnotesize
	\begin{axis}[%
		width=.8\textwidth,
		height=0.5\textwidth,
		at={(1.128in,0.894in)},
		scale only axis,
		xmin=0,
		xmax=6,
		xlabel={Number of users},
		xtick={0, 1, 2, 3, 4, 5, 6},
		xticklabels={$ $, $200$, $ 400 $, $ 600 $, $ 800 $, $ 1000 $},
            xlabel style={font=\footnotesize},
            ylabel style={font=\footnotesize},
            tick label style={font=\footnotesize},
            ymin= 400,
            ymax= 90000,
		ylabel = {Bandwidth (B)},
		ylabel shift=-5pt,
		yticklabel shift={0cm},
		axis background/.style={fill=white},
		legend columns=2,
		legend style={legend cell align=left, align=left, fill=none, draw=none,inner sep=-0pt, row sep=0pt, font = \tiny},
		legend pos = north west,
		%
		minor y tick num=5,
		]

        \addplot [color=E, solid, mark=asterisk, mark options={solid, E}]
		table[row sep=crcr]{%
            1 6600 \\
			2 13200 \\
			3 19800\\ 
			4 26400\\ 
			5 33000\\
		};
		\addlegendentry{\nano~(SH)}

       \addplot [color=A, solid, mark=diamond*, mark options={solid, A}]
		table[row sep=crcr]{%
                1 13233 \\ 
			2 26433\\ 
			3 39633\\
			4 52833\\
			5 66033\\
		};
		\addlegendentry{\nano~(MS)}

  
	\end{axis}

\end{tikzpicture}%
					\caption{Assisting node}\label{fig:exp:BWS:node}
			\end{subfigure}
			
		} 

	\caption{Outbound Bandwidth in the Setup phase}\label{fig:exp:BWS} 
	\vspace{-10pt}
\end{figure*}

\section{Evaluation} 
We analytically and experimentally evaluate and compare \nano with the state of the art~\cite{ccs/BellBGL020,popets/GuoPSBB24,sp/MaWAPR23}.\footnote{
Experimental analysis for resilience against different inference attacks is non-essential in secure-aggregation literature\mbox{~\cite{sp/MaWAPR23,ccs/BonawitzIKMMPRS17}}.
Such attacks against user gradients are ineffective since the masked gradients are essentially random values drawn uniformly from the corresponding distribution,
unlike gradient perturbation methods (\eg, differential privacy\mbox{~\cite{ccs/AbadiCGMMT016}}).
}

\subsection{Analytical Evaluation}\label{sec:analytical}
We compare the analytical performance of \nano and the state-of-the-art in Tables~\ref{tab:analytical_setup} and \ref{tab:analytical_agg}. 
Setting up \nano only incurs two runs of key generation protocol and $k$ runs of key agreement protocol.
Note that $k$ can be as small as~$2$, while the number of participating decryptors $|D'|$ is at least $64$ in Flamingo~\cite{sp/MaWAPR23}.

We also compare \nano with the maliciously-secure version of MicroSecAgg~\cite{popets/GuoPSBB24}.
MicroSecAgg incurs a linear number of signature verification, key agreement, secret sharing, and symmetric encryption and decryption algorithms, which make it more costly than \nano.
The key material of the protocol of Bell~\etal~\cite{ccs/BellBGL020} cannot be reused across multiple rounds (see Table~\ref{tab:analytical_setup}); therefore, the setup phase should be repeated in each aggregation phase.
In the Aggregation phase of \nano, the user only performs $k$ $\PRF$ invocations, $k + 1$ summation, and two signature generations to achieve malicious security.
Given that $k$ is significantly smaller than the number of participating users/decryptors, this results in higher efficiency than its counterparts.

\subsection{Experimental Evaluation}\label{sec:experiments}

\subsubsection{Implementation}
We implemented our protocol in Python with $1500$ lines of code.
We used \texttt{coincurve}~\cite{lib:coincurve} to implement public key primitives based on the elliptic curve (EC), including ECDSA signature and EC Diffie-Hellman (ECDH) key exchange protocols.
We used AES for $\PRF$ to generate masks.
To implement communication between all the parties, we used the standard Python socket.

\myPara{Parameter choice and counterpart comparison} 
As alluded to, a large body of research on privacy-preserving aggregation protocol for FL exists.
We have selected two most efficient works, namely, MicroSecAgg {\cite{popets/GuoPSBB24}} and Flamingo {\cite{sp/MaWAPR23}}, as our counterparts.
MicroSecAgg enables the reuse of masking terms in each iteration, achieving significant communication and computation efficiency.
In addition to having a one-time setup, Flamingo utilizes helper nodes (decryptors) to improve the communication and computation overhead.
The parameters for our protocol and its counterpart to achieve $128$-bit security are selected as follows.

\nano:
We used \texttt{secp256k1} curve with $256$-bit group order for ECDSA and ECDH protocols.
We set $k=3$ and evaluate our scheme in semi-honest (SH) and malicious (MS) settings.

MicroSecAgg~\cite{popets/GuoPSBB24}: 
We compare \nano with two MicroSecAgg instantiations,  MicroSecAgg$_1$ (MicroSecAgg$_{\text{DL}}$ in \cite{popets/GuoPSBB24}) and MicroSecAgg$_2$ (MicroSecAgg$_{\text{gDL}}$ in \cite{popets/GuoPSBB24}), both instantiated in the semi-honest setting. We note that the authors propose another instantiation, MicroSecAgg$_{\text{CL}}$, however they reported the same performance (see Figures 2 and 4 in \cite{popets/GuoPSBB24}) with MicroSecAgg$_{\text{gDL}}$, with no publicly available source code.
In MicroSecAgg$_1$,
all users communicate with each other via the aggregation server (star topology).
Following~\cite{ccs/BellBGL020}, in MicroSecAgg$_2$, users are grouped in sets of size $\log n$ to improve efficiency.
We selected the group size of $50$ in MicroSecAgg$_2$ to optimize for efficiency.
We used the suggested parameters and instantiations~\cite{popets/GuoPSBB24}, including $2048$-bit prime, XSalsa20 for authenticated encryption, Diffie-Hellman key exchange, and SHA-256 for hash function.
The discrete-logarithm-solving version of MicroSecAgg can handle weights up to $20$ bits.

Flamingo~\cite{sp/MaWAPR23}:
We compare \nano with two Flamingo instantiations, with $64$ and $128$ decryptors, which we call
Flamingo$_{64}$ and Flamingo$_{128}$.
Both instantiations are identical otherwise.
Regular users (\FlamingoUserFirst, \FlamingoUserSec) mask their values for each aggregation round.
The server then aggregates these masked values 
and removes the aggregated mask with the help of a designated set of user decryptors (\FlamingoDecFirst, \FlamingoDecSec).
Both configurations operate in a malicious setting, 
with all users (including decryptors) 
communicating via the server in a star topology.
We adopted the existing parameters~\cite{sp/MaWAPR23}, instantiated $\PRF$~with AES, used AES-GCM for authenticated encryption and SHA-256 as the hash function, the \texttt{p-256} curve for ECDSA signatures, and ElGamal over the \texttt{p-256} curve for asymmetric encryption.
We assume Flamingo users will remain online continuously throughout the protocol.
User dropping out would hurt Flamingo's performance.

\myPara{Evaluation setting and metrics}
We evaluated our protocol and its counterpart on a local Macbook Pro 2021 with a $3.2$ GHz M1 Pro CPU and $16$ GB RAM.

We considered a range of $200$ to $1,000$ users, three assisting nodes, and one aggregation server.
We measure the computation time and the outbound bandwidth cost of the server, the users, and the assisting nodes (in our protocol) in both the Setup and Aggregation phases.

To show the performance of \nano with high-dimensional gradient vectors, we report the performance with a weight vector $\vec{w}$ of $16,000$ gradients.
In fact, compared to the existing works (\eg,~\cite{popets/GuoPSBB24,sp/MaWAPR23}), our protocol offers a much higher efficiency for weight vectors with high dimensions since it does not require any expensive operations (\eg, solving for discrete logarithms or class-group operations) to compute the final model from local updates.

MicroSecAgg only reported performance for single gradients.
We projected their results for $16,000$ gradient vectors.
We also report the results for a single value gradient for a fair comparison.
Flamingo presented performance results for a vector of size $16,000$.
We ran all the experiments $10$ times and reported the average.

\begin{figure*}[!t]
	\centering
		\resizebox{1\textwidth}{!}{
			\begin{subfigure}{.3\textwidth}
%
%
\definecolor{A}{HTML}{e6194B}%
\definecolor{B}{HTML}{f58231}%
\definecolor{C}{HTML}{4363d8}%
\definecolor{D}{HTML}{911eb4}%
\definecolor{E}{HTML}{3cb44b}%
\definecolor{F}{rgb}{0.92900,0.69400,0.12500}%
\definecolor{G}{HTML}{808000}%
\definecolor{H}{HTML}{000000}%
\begin{tikzpicture}
	\begin{axis}[%
		width=.8\textwidth,
		height=0.5\textwidth,
		at={(1.128in,0.894in)},
		scale only axis,
		xmin=0,
		xmax=6,
		xlabel={Number of users},
		xtick={0, 1, 2, 3, 4, 5, 6},
		xticklabels={$ $, $200$, $ 400 $, $ 600 $, $ 800 $, $ 1000 $},
            xlabel style={font=\footnotesize},
            ylabel style={font=\footnotesize},
            tick label style={font=\footnotesize},
		ymax=10000000000000,
            ymode=log,
		ylabel = {Delay (ms) (log)},
		ylabel shift=-5pt,
		yticklabel shift={0cm},
		axis background/.style={fill=white},
		legend columns=2,
		legend style={legend cell align=left, align=left, fill=none, draw=none,inner sep=-0pt, row sep=0pt, font = \tiny},
		legend pos = north west,
		%
		minor y tick num=5,
		]
  
        \addplot [color=E, solid, mark=asterisk, mark options={solid, E}]
		table[row sep=crcr]{%
                1 14.451 \\ 
			2 24.1613 \\ 
			3 33.603 \\ 
			4 44.8899 \\ 
			5 55.273 \\ 
		};
		\addlegendentry{\nano~(SH)}

		\addplot [color=A, solid, mark=diamond*, mark options={solid, A}]
		table[row sep=crcr]{%
			1 71.3816\\ 
			2 83.0221\\ 
			3 115.9539\\ 
			4 142.1985\\ 
			5 309.8507\\ 
		};
		\addlegendentry{\nano~(MS)}
		
		\addplot [color=B, dashed, mark=square*, mark options={solid, B}]
		table[row sep=crcr]{%
			1 26915200\\ 
			2 39625600\\ 
			3 53830400\\ 
			4 75513600\\ 
			5 101681600\\ 
		};
		\addlegendentry{MicroSecAgg$_1$}

		\addplot [color=C, loosely dotted, mark=triangle*, mark options={solid, C}]
		table[row sep=crcr]{%
			1 31846400\\ 
			2 41294400\\ 
			3 52265600\\ 
			4 63694400\\ 
			5 74056000\\ 
		};
		\addlegendentry{MicroSecAgg$_2$(50)}

		\addplot [color=H, dashed, mark=pentagon*, mark options={solid, H}]
		table[row sep=crcr]{%
			1 3617.555\\ %
			2 4943.7165\\ %
			3 7378.309\\ %
			4 10060.772\\ %
			5 14610.788\\ 
		};
		\addlegendentry{\FlamingoServerFirst}

		\addplot [color=D, loosely dotted, mark=otimes*, mark options={solid, D}]
		table[row sep=crcr]{%
			1 4990.282\\ %
			2 8532.758\\ %
			3 11814.192\\ %
			4 19848.084\\ %
			5 27575.625\\ 
		};
		\addlegendentry{\FlamingoServerSec}

    \draw[solid, black, latex-latex, line width=0.01pt] (515, 3.7912463990471709) -- (515, 9.784907657724967);
    \node[solid, black, latex-latex] at (548, 6.8380770284) {\tiny 264x};

    \draw[solid, black, latex-latex, line width=0.01pt] (515, 9.784907657724967) -- (515,18.684907657724967);
    \node[solid, black, latex-latex] at (551, 13.5380770284) {\tiny 6959x};

 \end{axis}

\end{tikzpicture}%
					\caption{Server}\label{fig:exp:IM:server} 
			\end{subfigure}
			\begin{subfigure}{.3\textwidth}
%
%
\definecolor{A}{HTML}{e6194B}%
\definecolor{B}{HTML}{f58231}%
\definecolor{C}{HTML}{4363d8}%
\definecolor{D}{HTML}{911eb4}%
\definecolor{E}{HTML}{3cb44b}%
\definecolor{F}{rgb}{0.92900,0.69400,0.12500}%
\definecolor{G}{HTML}{808000}%
\definecolor{H}{HTML}{000000}%
\begin{tikzpicture}
\footnotesize
	\begin{axis}[%
		width=.8\textwidth,
		height=0.5\textwidth,
		at={(1.128in,0.894in)},
		scale only axis,
		xmin=0,
		xmax=6,
		xlabel={Number of users},
		xtick={0, 1, 2, 3, 4, 5, 6},
		xticklabels={$ $, $200$, $ 400 $, $ 600 $, $ 800 $, $ 1000 $},
            xlabel style={font=\footnotesize},
            ylabel style={font=\footnotesize},
            tick label style={font=\footnotesize},        
		ymax=1000000000000,
            ymode=log,
		ylabel = {Delay (ms) (log)},
		ylabel shift=-5pt,
		yticklabel shift={0cm},
		axis background/.style={fill=white},
		legend columns=2,
		legend style={legend cell align=left, align=left, fill=none, draw=none,inner sep=-0pt, row sep=0pt, font = \tiny},
		legend pos = north west,
		%
		]
  
        \addplot [color=E, solid, mark=asterisk, mark options={solid, E}]
		table[row sep=crcr]{%
                1 2.0726 \\ 
			2 2.0726 \\
			3 2.0726\\ 
			4 2.0726 \\
			5 2.0726 \\ 
		};
		\addlegendentry{\nano~(SH)}
		\addplot [color=A, solid, mark=diamond*, mark options={solid, A}]
		table[row sep=crcr]{%
                1 4.8049 \\ 
			2 4.8049\\ 
			3 4.8049\\
			4 4.8049\\
			5 4.8049\\
		};
		\addlegendentry{\nano~(MS)}	      
  
  		\addplot [color=B, dashed, mark=square*, mark options={solid, B}]
		table[row sep=crcr]{%
			1 251264\\ 
			2 254768\\ 
			3 258096\\ 
			4 262080\\ 
			5 266336\\ 
		};
		\addlegendentry{MicroSecAgg$_1$}

            \addplot [color=C, loosely dotted, mark=triangle*, mark options={solid, C}]
		table[row sep=crcr]{%
			1 248848\\ 
			2 250128\\ 
			3 251088\\ 
			4 253280\\ 
			5 253920\\ 
		};
		\addlegendentry{MicroSecAgg$_2$(50)}
    
		\addplot [color=H, dashed, mark=pentagon*, mark options={solid, H}]
		table[row sep=crcr]{%
			1 136.69421\\ %
			2 140.282807\\ %
			3 148.747371\\ %
			4 152.341664\\ %
			5 156.361156\\ 
		};
		\addlegendentry{\FlamingoUserFirst}

		\addplot [color=D, loosely dotted, mark=otimes*, mark options={solid, D}]
		table[row sep=crcr]{%
			1 197.827737\\ %
			2 222.71212\\ %
			3 234.354903\\ %
			4 240.58783\\ %
			5 246.377638\\ 
		};
		\addlegendentry{\FlamingoUserSec}

		\addplot [color=teal, dashed, mark=halfcircle*, mark options={solid, teal}]
		table[row sep=crcr]{%
			1 206.735646\\ 
			2 249.001978\\ 
			3 348.57994\\ 
			4 500.548148\\ 
			5 567.10609\\ 
		};
		\addlegendentry{\FlamingoDecFirst}

		\addplot [color=magenta, loosely dotted, mark=halfsquare*, mark options={solid, magenta}]
		table[row sep=crcr]{%
			1 281.677259\\ 
			2 378.041642\\ 
			3 485.998933\\ 
			4 670.12304\\ 
			5 819.937184\\ 
		};
		\addlegendentry{\FlamingoDecSec}
  
    \draw[solid, black, latex-latex, line width=0.01pt] (515, 0.6912463990471709) -- (515,5.184907657724967);
    \node[solid, black, latex-latex] at (545, 2.7380770284) {\tiny 75x};

    \draw[solid, black, latex-latex, line width=0.01pt] (515, 5.184907657724967) -- (515,12.584907657724967);
    \node[solid, black, latex-latex] at (552, 8.5380770284) {\tiny 1703x};

	\end{axis}

%
 
\end{tikzpicture}%
					\caption{User} \label{fig:exp:IM:user} 
			\end{subfigure}
			\begin{subfigure}{.3\textwidth}
%
%
\definecolor{A}{HTML}{e6194B}%
\definecolor{B}{HTML}{f58231}%
\definecolor{C}{HTML}{4363d8}%
\definecolor{D}{HTML}{911eb4}%
\definecolor{E}{HTML}{3cb44b}%
\definecolor{F}{rgb}{0.92900,0.69400,0.12500}%
\definecolor{G}{HTML}{808000}%
\definecolor{H}{HTML}{000000}%
\begin{tikzpicture}
\footnotesize
	\begin{axis}[%
		width=.8\textwidth,
		height=0.5\textwidth,
		at={(1.128in,0.894in)},
		scale only axis,
		xmin=0,
		xmax=6,
		xlabel={Number of users},
		xtick={0, 1, 2, 3, 4, 5, 6},
		xticklabels={$ $, $200$, $ 400 $, $ 600 $, $ 800 $, $ 1000 $},
            xlabel style={font=\footnotesize},
            ylabel style={font=\footnotesize},
            tick label style={font=\footnotesize},
		ymin= 70,
		ymax= 900,
		ylabel = {Delay (ms)},
		ylabel shift=-5pt,
		yticklabel shift={0cm},
		axis background/.style={fill=white},
		legend columns=2,
		legend style={legend cell align=left, align=left, fill=none, draw=none,inner sep=-0pt, row sep=0pt, font = \tiny},
		legend pos = north west,
		%
		]

        \addplot [color=E, solid, mark=asterisk, mark options={solid, E}]
		table[row sep=crcr]{%
                1 136.23 \\ 
			2 270.06 \\ 
			3 409.04\\ 
			4 544.65\\ 
			5 681.81\\ 
		};
		\addlegendentry{\nano~(SH)}
  
       \addplot [color=A, solid, mark=diamond*, mark options={solid, A}]
		table[row sep=crcr]{%
			1 138.53\\ 
			2 274.275\\ 
			3 415.7841 \\ 
			4 577.8709\\ 
			5 721.9098\\ 
		};
		\addlegendentry{\nano~(MS)}

	\end{axis}
 
\end{tikzpicture}%
					\caption{Assisting node} \label{fig:exp:IM:node}
			\end{subfigure}
		} 
 
	\caption{Computation in the Aggregation phase}\label{fig:exp:IM} 
	 \vspace{-12pt}
\end{figure*}

\begin{figure*}[!t]
	\centering
		\resizebox{1\textwidth}{!}{
			\begin{subfigure}{.3\textwidth}
%
%
\definecolor{A}{HTML}{e6194B}%
\definecolor{B}{HTML}{f58231}%
\definecolor{C}{HTML}{4363d8}%
\definecolor{D}{HTML}{911eb4}%
\definecolor{E}{HTML}{3cb44b}%
\definecolor{F}{rgb}{0.92900,0.69400,0.12500}%
\definecolor{G}{HTML}{808000}%
\definecolor{H}{HTML}{000000}%
\begin{tikzpicture}
	\footnotesize
	\begin{axis}[%
		width=.8\textwidth,
		height=0.5\textwidth,
		at={(1.128in,0.894in)},
		scale only axis,
		xmin=0,
		xmax=6,
		xlabel={Number of users},
		xtick={0, 1, 2, 3, 4, 5, 6},
		xticklabels={$ $, $200$, $ 400 $, $ 600 $, $ 800 $, $ 1000 $},
            xlabel style={font=\footnotesize},
            ylabel style={font=\footnotesize},
            tick label style={font=\footnotesize},  
		ymax=1000000000000000,
            ymode=log,
		ylabel = {Bandwidth (B) (log)},
		ylabel shift=-5pt,
		yticklabel shift={0cm},
		axis background/.style={fill=white},
		legend columns=2,
		legend style={legend cell align=left, align=left, fill=none, draw=none,inner sep=-0pt, row sep=0pt, font = \tiny},
		legend pos = north west,
		%
		minor y tick num=5,
		]

        \addplot [color=B, dashed, mark=square*, mark options={solid, B}]
		table[row sep=crcr]{%
			1 32820512819.2\\ 
			2 98461538460.8\\ 
			3 229743589742.4\\ 
			4 410256410256\\ 
			5 640000000000\\ 
		};
		\addlegendentry{MicroSecAgg$_1$}

            \addplot [color=C, loosely dotted, mark=triangle*, mark options={solid, C}]
		table[row sep=crcr]{%
			1 32820512819.2\\ 
			2 98461538460.8\\ 
			3 229743589742.4\\ 
			4 410256410256\\ 
			5 640000000000\\ 
		};
		\addlegendentry{MicroSecAgg$_2$(50)}

		\addplot [color=H, dashed, mark=pentagon*, mark options={solid, H}]
		table[row sep=crcr]{%
			1 39273248\\ 
			2 78288448\\ 
			3 117303648\\ 
			4 156318848\\ 
			5 195334048\\ 
		};
		\addlegendentry{\FlamingoServerFirst}

		\addplot [color=D, loosely dotted, mark=otimes*, mark options={solid, D}]
		table[row sep=crcr]{%
			1 118698784\\ 
			2 236357184\\ 
			3 354015584\\ 
			4 471673984\\ 
			5 589332384\\ 
		};
		\addlegendentry{\FlamingoServerSec}

		\addplot [color=A, solid, mark=diamond*, mark options={solid, A}]
		table[row sep=crcr]{%
                1 12800064 \\ 
			2 25600064 \\ %
			3 38400064\\ %
			4 51200064 \\ %
			5 64000064 \\ %
		};
		\addlegendentry{\nano~(MS)}
  
          \addplot [color=E, solid, mark=asterisk, mark options={solid, E}]
		table[row sep=crcr]{%
                1 12800000 \\ 
			2 25600000 \\ %
			3 38400000\\ %
			4 51200000 \\ %
			5 64000000 \\ %
		};
		\addlegendentry{\nano~(SH)}
  80,000,000

        \draw[solid, black, latex-latex, line width=0.01pt] (515, 17.784907657724967) -- (515,19.184907657724967);
        \node[solid, black, latex-latex] at (543, 18.4449076577) {\tiny 3x};
    
        \draw[solid, black, latex-latex, line width=0.01pt] (515, 19.184907657724967) -- (515,27.384907657724967);
        \node[solid, black, latex-latex] at (555, 23.0349076577) {\tiny 3276x};
      
	\end{axis}

\end{tikzpicture}%
					\caption{Server}\label{fig:exp:comm:server}
			\end{subfigure}
			\begin{subfigure}{.3\textwidth}
%
%
\definecolor{A}{HTML}{e6194B}%
\definecolor{B}{HTML}{f58231}%
\definecolor{C}{HTML}{4363d8}%
\definecolor{D}{HTML}{911eb4}%
\definecolor{E}{HTML}{3cb44b}%
\definecolor{F}{rgb}{0.92900,0.69400,0.12500}%
\definecolor{G}{HTML}{808000}%
\definecolor{H}{HTML}{000000}%
\begin{tikzpicture}
	\footnotesize
	\begin{axis}[%
		width=.8\textwidth,
		height=0.5\textwidth,
		at={(1.128in,0.894in)},
		scale only axis,
		xmin=0,
		xmax=6,
		xlabel={Number of users},
		xtick={0, 1, 2, 3, 4, 5, 6},
		xticklabels={$ $, $200$, $ 400 $, $ 600 $, $ 800 $, $ 1000 $},
            xlabel style={font=\footnotesize},
            ylabel style={font=\footnotesize},
            tick label style={font=\footnotesize},  
		ymax= 100000000000,
            ymode=log,
		ylabel = {Bandwidth (B) (log)},
		ylabel shift=-5pt,
		yticklabel shift={0cm},
		axis background/.style={fill=white},
		legend columns=2,
		legend style={legend cell align=left, align=left, fill=none, draw=none,inner sep=-0pt, row sep=0pt, font = \tiny},
		legend pos = north west,
		%
		minor y tick num=5,
		]

		\addplot [color=A, solid, mark=diamond*, mark options={solid, A}]
		table[row sep=crcr]{%
                1 64272 \\ 
			2 64272\\ 
			3 64272\\
			4 64272\\
			5 64272\\
		};
		\addlegendentry{\nano~(MS)}		

        \addplot [color=E, solid, mark=asterisk, mark options={solid, E}]
		table[row sep=crcr]{%
                1 64016 \\ 
			2 64016 \\ 
			3 64016 \\ 
			4 64016 \\
			5 64016 \\
		};
		\addlegendentry{\nano~(SH)}
  
            \addplot [color=B, dashed, mark=square*, mark options={solid, B}]
		table[row sep=crcr]{%
			1 17858064\\ 
			2 17858064\\ 
			3 17858064\\ 
			4 17858064\\ 
			5 17858064\\ 
		};
		\addlegendentry{MicroSecAgg$_1$}

            \addplot [color=C, loosely dotted, mark=triangle*, mark options={solid, C}]
		table[row sep=crcr]{%
			1 17858064\\ 
			2 17858064\\ 
			3 17858064\\ 
			4 17858064\\ 
			5 17858064\\ 
		};
		\addlegendentry{MicroSecAgg$_2$(50)}

		\addplot [color=D, loosely dotted, mark=otimes*, mark options={solid, D}]
		table[row sep=crcr]{%
			1 135168\\ %
			2 135552\\ %
			3 135552\\ %
			4 135936\\ %
			5 135936\\ 
		};
		\addlegendentry{\FlamingoUserSec}

		\addplot [color=H, dashed, mark=pentagon*, mark options={solid, H}]
		table[row sep=crcr]{%
			1 133120\\ %
			2 133504\\ %
			3 133504\\ %
			4 133888\\ %
			5 133888\\ 
		};
		\addlegendentry{\FlamingoUserFirst}

		\addplot [color=teal, dashed, mark=halfcircle*, mark options={solid, teal}]
		table[row sep=crcr]{%
			1 145984\\ 
			2 159168\\ 
			3 171968\\ 
			4 185152\\ 
			5 197952\\ 
		};
		\addlegendentry{\FlamingoDecFirst}

		\addplot [color=magenta, loosely dotted, mark=halfsquare*, mark options={solid, magenta}]
		table[row sep=crcr]{%
			1 148032\\ 
			2 161216\\ 
			3 174016\\ 
			4 187200\\ 
			5 200000\\ 
		};
		\addlegendentry{\FlamingoDecSec}

    \draw[solid, black, latex-latex, line width=0.01pt] (515, 10.9912463990471709) -- (515,11.834907657724967);
    \node[solid, black, latex-latex] at (539, 11.4130770284) {\tiny 2x};

    \draw[solid, black, latex-latex, line width=0.01pt] (515, 11.834907657724967) -- (515,16.784907657724967);
    \node[solid, black, latex-latex] at (544, 13.7849076577) {\tiny 133x};

 	\end{axis}

\end{tikzpicture}%
					\caption{User}\label{fig:exp:comm:user}
			\end{subfigure}
			\begin{subfigure}{.3\textwidth}
%
%
\definecolor{A}{HTML}{e6194B}%
\definecolor{B}{HTML}{f58231}%
\definecolor{C}{HTML}{4363d8}%
\definecolor{D}{HTML}{911eb4}%
\definecolor{E}{HTML}{3cb44b}%
\definecolor{F}{rgb}{0.92900,0.69400,0.12500}%
\definecolor{G}{HTML}{808000}%
\definecolor{H}{HTML}{000000}%
\begin{tikzpicture}
	\footnotesize
	\begin{axis}[%
		width=.8\textwidth,
		height=0.5\textwidth,
		at={(1.128in,0.894in)},
		scale only axis,
		xmin=0,
		xmax=6,
		xlabel={Number of users},
		xtick={0, 1, 2, 3, 4, 5, 6},
		xticklabels={$ $, $200$, $ 400 $, $ 600 $, $ 800 $, $ 1000 $},
            xlabel style={font=\footnotesize},
            ylabel style={font=\footnotesize},
            tick label style={font=\footnotesize},  
            ymin= 63500,
		ymax= 64700,
		ylabel = {Bandwidth (B)},
		ylabel shift=-5pt,
		yticklabel shift={0cm},
		axis background/.style={fill=white},
		legend columns=2,
		legend style={legend cell align=left, align=left, fill=none, draw=none,inner sep=-0pt, row sep=0pt, font = \tiny},
		legend pos = north west,
		%
		]
        \addplot [color=E, solid, mark=asterisk, mark options={solid, E}]
		table[row sep=crcr]{%
            1 64008 \\ 
			2 64008 \\
			3 64008\\ 
			4 64008\\ 
			5 64008\\
		};
		\addlegendentry{\nano~(SH)}
  
       \addplot [color=A, solid, mark=diamond*, mark options={solid, A}]
		table[row sep=crcr]{%
            1 64072 \\ 
			2 64072\\ 
			3 64072\\
			4 64072\\
			5 64072\\
		};
		\addlegendentry{\nano~(MS)}


\end{axis}

\end{tikzpicture}%
					\caption{{Assisting node}}\label{fig:exp:comm:node}
			\end{subfigure}
		} 

	\caption{Outbound Bandwidth in the Aggregation phase }\label{fig:exp:BW} 
 
\end{figure*}

%


\definecolor{A}{HTML}{e6194B}
\definecolor{B}{HTML}{f58231}
\definecolor{C}{HTML}{4363d8}
\definecolor{D}{HTML}{911eb4}
\definecolor{E}{HTML}{3cb44b}
\definecolor{F}{HTML}{eeb220}
\definecolor{G}{HTML}{808000}
\definecolor{H}{HTML}{000000}

 \begin{figure}[!t] 
	\centering
		\pgfplotslegendfromname{legend_aggregation}

			\hspace{-1em}
				\begin{subfigure}{.49\columnwidth}
%
%
\definecolor{A}{HTML}{e6194B}%
\definecolor{B}{HTML}{f58231}%
\definecolor{C}{HTML}{4363d8}%
\definecolor{D}{HTML}{911eb4}%
\definecolor{E}{HTML}{3cb44b}%
\definecolor{F}{rgb}{0.92900,0.69400,0.12500}%
\definecolor{G}{HTML}{808000}%
\definecolor{H}{HTML}{000000}%
\begin{tikzpicture}
	\begin{axis}[%
		width=0.8\textwidth,
		height=0.4\textwidth,
		at={(1.128in,0.894in)},
		scale only axis,
		xmin=0,
		xmax=6,
		xlabel={Number of users},
		xtick={0, 1, 2, 3, 4, 5, 6},
            xticklabels={$ $, $200$, $ 400 $, $ 600 $, $ 800 $, $ 1000 $},
            xlabel style={font=\footnotesize},
            ylabel style={font=\footnotesize},
            tick label style={font=\footnotesize},
		ymax=1000000,
            ymode=log,
		ylabel = {Delay (ms)},
		ylabel shift=-5pt,
		yticklabel shift={0cm},
		axis background/.style={fill=white},
            legend to name=legend_setup,
		legend columns=3,
		legend style={font=\tiny, draw=black, fill=none},
		%
		minor y tick num=5,
		]

        \addplot [color=E, solid, mark=asterisk, mark options={solid, E}]
		table[row sep=crcr]{%
                1 33.21 \\ 
			2 63.21 \\ 
			3 93.57 \\ 
			4 123.96 \\ 
			5 152.82 \\ 
		};
		\addlegendentry{\nano~(SH)}

		\addplot [color=A, solid, mark=diamond*, mark options={solid, A}]
		table[row sep=crcr]{%
			1 33.46\\ 
			2 63.59\\ 
			3 95.78\\ 
			4 126.35\\ 
			5 155.78\\ 
		};
		\addlegendentry{\nano~(MS)}
		
		\addplot [color=B, dashed, mark=square*, mark options={solid, B}]
		table[row sep=crcr]{%
			1 98.88\\ 
			2 396.92\\ 
			3 978.37\\ 
			4 1826.53\\ 
			5 2885.09\\ 
		};
		\addlegendentry{MicroSecAgg$_1$}

		\addplot [color=C, loosely dotted, mark=triangle*, mark options={solid, C}]
		table[row sep=crcr]{%
			1 54.93\\ 
			2 85.23\\ 
			3 123.61\\ 
			4 166.04\\ 
			5 210.48\\ 
		};
		\addlegendentry{MicroSecAgg$_2$(50)}

		\addplot [color=H, dashed, mark=pentagon*, mark options={solid, H}]
		table[row sep=crcr]{%
			1 36720.64\\ %
			2 36720.64\\ %
			3 36720.64\\ %
			4 36720.64\\ %
			5 36720.64\\ 
		};
		\addlegendentry{\FlamingoServerSec}

		\addplot [color=D, loosely dotted, mark=otimes*, mark options={solid, D}]
		table[row sep=crcr]{%
			1 302490.93\\ %
			2 302490.93\\ %
			3 302490.93\\ %
			4 302490.93\\ %
			5 302490.93\\ 
		};
		\addlegendentry{\FlamingoServerSec}

	\end{axis}

\end{tikzpicture}%
					\caption{{Setup Phase}}\label{fig:exp:anc:total:setup}
			\end{subfigure}\hspace{2mm}
			\begin{subfigure}{.49\columnwidth}
%
%
\definecolor{A}{HTML}{e6194B}%
\definecolor{B}{HTML}{f58231}%
\definecolor{C}{HTML}{4363d8}%
\definecolor{D}{HTML}{911eb4}%
\definecolor{E}{HTML}{3cb44b}%
\definecolor{F}{rgb}{0.92900,0.69400,0.12500}%
\definecolor{G}{HTML}{808000}%
\definecolor{H}{HTML}{000000}%
\begin{tikzpicture}
\small	
	\begin{axis}[%
		width=0.8\textwidth,
		height=0.4\textwidth,
		at={(1.128in,0.894in)},
		scale only axis,
		xmin=0,
		xmax=6,
		xlabel={Number of users},
		xtick={0, 1, 2, 3, 4, 5, 6},
            xticklabels={$ $, $200$, $ 400 $, $ 600 $, $ 800 $, $ 1000 $},
            xlabel style={font=\footnotesize},
            ylabel style={font=\footnotesize},
            tick label style={font=\footnotesize},            
		ymax=1000000000,
            ymode=log,
		ylabel shift=-5pt,
		yticklabel shift={0cm},
		axis background/.style={fill=white},
            legend to name=legend_aggregation,
		legend columns=3,
		legend style={font=\tiny, draw=black, fill=none},
		%
		minor y tick num=5,
		]

        \addplot [color=E, solid, mark=asterisk, mark options={solid, E}]
		table[row sep=crcr]{%
                1 425.2136 \\ 
			2 836.4139 \\ 
			3 1262.7956 \\ 
			4 1680.9125 \\ 
			5 2102.7756 \\ 
		};
		\addlegendentry{\nano~(SH)}

		\addplot [color=A, solid, mark=diamond*, mark options={solid, A}]
		table[row sep=crcr]{%
			1 491.7765\\ 
			2 910.652\\ 
			3 1368.1111\\ 
			4 1880.6161\\ 
			5 2480.385\\ 
		};
		\addlegendentry{\nano~(MS)}
		
		\addplot [color=B, dashed, mark=square*, mark options={solid, B}]
		table[row sep=crcr]{%
			1 27166464\\ 
			2 39880368\\ 
			3 54088496\\ 
			4 75775680\\ 
			5 101947936\\ 
		};
		\addlegendentry{MicroSecAgg$_1$}

		\addplot [color=C, loosely dotted, mark=triangle*, mark options={solid, C}]
		table[row sep=crcr]{%
			1 32095248\\ 
			2 41544528\\ 
			3 52516688\\ 
			4 63947680\\ 
			5 74309920\\ 
		};
		\addlegendentry{MicroSecAgg$_2$(50)}

		\addplot [color=H, dashed, mark=pentagon*, mark options={solid, H}]
		table[row sep=crcr]{%
			1 8236.901114\\ %
			2 12042.026251\\ %
			3 20316.340787\\ %
			4 32498.32864\\ %
			5 41054.824932\\ 
		};
		\addlegendentry{\FlamingoServerFirst}

		\addplot [color=D, loosely dotted, mark=otimes*, mark options={solid, D}]
		table[row sep=crcr]{%
			1 15920.848553\\ %
			2 28637.648936\\ %
			3 44258.982743\\ %
			4 75069.17871\\ %
			5 101237.624526\\ 
		};
		\addlegendentry{\FlamingoServerSec}

	\end{axis}

\end{tikzpicture}%
					\caption{{Aggregation Phase}}\label{fig:exp:anc:total:agg}
			\end{subfigure}

	\caption{{Overall Computation Overhead of Setup and Aggregation}}\label{fig:exp:anc:total} 
	
\end{figure}

\subsubsection{Results}
 
\myPara{Setup phase: Computation} 
Figure~\ref{fig:exp:IMS} compares the computation cost of \nano in Setup with the ones in MicroSecAgg and Flamingo.
As depicted in Figure~\ref{fig:exp:IMS}, \nano outperforms MicroSecAgg for server and user computation in the semi-honest setting.
Server computation in Flamingo$_{64}$ is slightly better than the one in \nano ($0.15$ vs.~$0.19$ ms).
However, \nano outperforms Flamingo$_{128}$ server computation and users (including regular and decryptor) computation in both Flamingo$_{64}$ and Flamingo$_{128}$.

Specifically, 
in the semi-honest setting, our server does not incur any computation,
while in the malicious setting, the server takes $0.19$ ms to setup (generate an ECDSA key pair), compared to $376$ and $178$ ms for MicroSecAgg$_1$ and MicroSecAgg$_2$, and $0.15$ ms and $0.32$ ms for Flamingo$_{64}$ and Flamingo$_{128}$, respectively (Figure~\ref{fig:exp:IMS:server}).
In MicroSecAgg protocols, the server communicates all the keys among users.
Thus, our malicious scheme is $53$--$1,977\times$ (for the defined number of users range) faster than MicroSecAgg$_1$ and $117$--$936\times$ faster than MicroSecAgg$_2$.
However, in Flamingo, the server does not generate the required secret keys for users, rendering it very efficient.
Our malicious scheme is nearly $2\times$ faster than Flamingo$_{128}$ (for $200$ to $1,000$ users).

Figure~\ref{fig:exp:IMS:user} shows that the user computation in the semi-honest (resp.~malicious) \nano is $470$--$13,284\times$ (resp.~$352$--$9,962\times$) faster than MicroSecAgg$_1$ and $173\times$ (resp.~$129\times$) faster than MicroSecAgg$_2$, 
where each user only takes $0.18$ ms (resp.~$0.25$ ms) to setup, while MicroSecAgg$_1$ and MicroSecAgg$_2$ take $2,509$ ms and $33$ ms, respectively, for $1,000$ users.
Moreover, in the semi-honest (resp.~malicious) setting, \nano user is $142\times$ (resp.~$106\times$) faster than Flamingo$_{64}$ regular user and $283\times$ (resp.~$212\times$) faster than Flamingo$_{128}$ regular user.
\nano user is also $3,177\times$ (resp.~$2,382\times$) faster than Flamingo$_{64}$ decryptor and $12,791\times$ (resp.~$9,592\times$) faster than Flamingo$_{128}$ decryptor (for the specified user range).
Specifically, 
the regular user requires {$27$} ms and {$53$} ms for setup, 
while 
the decryptor takes {$600$} ms and {2416} ms 
for Flamingo$_{64}$ and Flamingo$_{128}$, 
respectively.

Similar to the server cost, \nano and Flamingo incur a constant user time regardless of the number of users.
Each user only needs to compute a single ECDH key pair (plus a single ECDSA key pair in the malicious setting), and compute the common secret with each assisting node.
In MicroSecAgg, however, after creating the DH keys and computing the common secret key with all other users, the user needs to sample masking term, secret share it to $n$ shares, and encrypt each of the $n$ shares with the common secret key of all the users in the system.
In Flamingo, each regular user is responsible for sampling the set of users designated as decryptors and verifying the signatures of public keys signed by decryptors.
However, in addition to its regular user responsibilities, each decryptor performs verifiable secret sharing, generates public keys for regular users, and signs and shares them with each user via the server.

Unlike MicroSecAgg and Flamingo,
\nano requires assisting nodes to aid secure aggregation.
However, the computation costs associated with assisting node in \nano is highly efficient, where it only takes $11$ ms and $51$ ms in semi-honest setting (and $11$ and $52$ ms in the malicious setting) to setup key materials for $200$ to $1,000$ users (Figure~\ref{fig:exp:IMS:node}).
Each node computes an ECDH key pair (plus an ECDSA key pair in the malicious setting) and a common secret seed for each user, and thus the overhead grows linearly w.r.t the number of users.

\myPara{Setup phase: Outbound bandwidth} 
We present the outbound bandwidth cost of all entities in \nano in Figure~\ref{fig:exp:BWS}.
In the malicious setting, for $1,000$ users (Figure~\ref{fig:exp:BWS:server}), the server transmits {\textasciitilden4--5} orders of magnitude 
less than MicroSecAgg$_1$
and {\textasciitilden3--4} orders of magnitude less than MicroSecAgg$_2$ (\ie, $33,000$ B vs.~$1,008$ MB in MicroSecAgg$_1$ and $240$ MB in MicroSecAgg$_2$).
Additionally, the server in \nano transmits {\textasciitilden2--3} orders of magnitude less than Flamingo$_{64}$ and {\textasciitilden3--4} orders of magnitude less than Flamingo$_{128}$ (\ie, $33,000$ B vs.~$15$ MB in Flamingo$_{64}$ and $49$ MB in Flamingo$_{128}$).
The server bandwidth in our malicious \nano protocol grows linearly as users increase because the server transmits its ECDSA public key to $200$ to $1,000$ users.
We note this is a one-time overhead, and with the assumption of a broadcast channel, this overhead could be a single message. 
Meanwhile, our semi-honest protocol does not require the server to transmit anything during the setup phase.
As shown in Figure~\ref{fig:exp:BWS:user}, user bandwidth in the semi-honest (resp.~malicious) \nano is $1,877$--$9,369\times$ (resp.~$804$--$4015\times$) less than MicroSecAgg.
Furthermore, 
it is $20$--$36\times$ (resp.~$9$--$16\times$) less than decryptor bandwidth in Flamingo.
Specifically, in \nano,
each user only transmits a fixed message of size $99$ B (resp.~$231$~B), while for MicroSecAgg$_1$ and MicroSecAgg$_2$ protocols it is approximately $905$ and $183$ KB, respectively.
For Flamingo$_{64}$ and Flamingo$_{128}$, it is about $2,013$ B and $3,612$ B, respectively.
This is because each user only needs to transmit an ECDH public key (plus an ECDSA public key in the malicious setting) to a fixed number of assisting nodes.

In MicroSecAgg, the users send (via the aggregation server) their DH public keys and the encrypted masking shares to all other users in the system.
In Flamingo, regular users do not send anything during the setup phase.
However, each decryptor transmits secret shares to other decryptors and sends public keys to all regular users via the server.
The bandwidth cost of assisting nodes in \nano is negligible, where each node only transmits to $200$--$1,000$ users a total of $6$--$32$ KB for the semi-honest protocol and $13$--$64$ KB message for the malicious protocol (Figure~\ref{fig:exp:BWS:node}).
The overhead is due to the broadcast of an ECDH public key (and an ECDSA public key in the malicious setting) to all the users (and the aggregating server), and thus, its bandwidth cost grows linearly w.r.t the number of users.

\myPara{Aggregation phase: Computation} 
Figure~\ref{fig:exp:IM} shows the computation cost in the aggregation phase for \nano, MicroSecAgg, and Flamingo.
In our protocol, all entities incur a small computation overhead in the Aggregation phase.
In particular, for a single gradient, the server computation time in our semi-honest (resp.~malicious) protocol is $815$--$1224\times$ (resp.~$139$--$187\times$) faster than MicroSecAgg$_1$ and $891$--$1,065\times$ (resp.~$117\times$--$222\times$) faster than MicroSecAgg$_2$.

When we consider the $16,000$ gradients vector, the server computation time in both of our semi-honest and malicious protocols is \textasciitilden5--6 orders of magnitude faster than those in MicroSecAgg$_1$ and MicroSecAgg$_2$.
Furthermore, the server computation time in our semi-honest (resp.~malicious) protocol is $264$--$499\times$ (resp.~$205$--$386\times$) faster than that in Flamingo$_{64}$ and Flamingo$_{128}$.
When there are $200$ users, the server computation time in semi-honest and malicious \nano is only $14$ ms and $71$ ms, respectively.
The server in MicroSecAgg protocols is \textasciitilden8 orders of magnitude slower, as it needs to solve small instances of the discrete logarithm problem.
Our server computation time in both settings grows linearly as the number of users increases.
In our semi-honest protocol, most of the aggregation server's cost is due to the linear summation of the weights and the aggregated masking terms provided by the assisting nodes.
The added $82$--$85\%$ of extra overhead in the malicious setting is due to the signature verification on the local weights sent by the parties and the signatures on the aggregated masks sent by the assisting nodes.

For a single gradient, user computation in semi-honest (resp.~malicious) \nano is $300$--$318\times$ (resp.~120$\times$--127$\times$) faster than the one in MicroSecAgg$_1$.
As shown in Figure~\ref{fig:exp:IM:user}, for $16,000$ gradients vector, the gap increases and the user computation in semi-honest and malicious \nano becomes \textasciitilden5 orders of magnitude faster than the ones in MicroSecAgg$_1$ and MicroSecAgg$_2$.
Additionally, for the same gradient vector size, the user computation time in both semi-honest and malicious \nano is is $33$--$119\times$ faster than the regular user in Flamingo$_{64}$ and Flamingo$_{128}$, as well as $118$--$396\times$ faster than the decryptor in Flamingo$_{64}$ and Flamingo$_{128}$.
More precisely, for $1,000$ users, the user computation time in our semi-honest (resp.~malicious) protocol is only $2.1$ ms (resp.~$4.8$ ms).
The user computation time of our protocol in both settings remains constant as the number of users grows.
Each user computes the masked update (plus two ECDSA signatures in the malicious setting).
Therefore, the cost is linear to the (small) number of assisting nodes rather than the number of users.

In MicroSecAgg, the user must compute the aggregation of the masking shares in the exponent after computing the mask update via additive masking.
In Flamingo, regular users only compute the mask update, and the decryptor assists the server in unmasking the aggregated result.
The computation time of the assisting nodes increases with the number of users, as shown in Figure~\ref{fig:exp:IM:node}; however, the overhead is small.
Specifically, for $200$--$1,000$ users the assisting node's time is $136$--$681$ ms in the semi-honest setting and $128$--$722$ ms in the malicious setting,
due to computing the aggregated masks for all the users ($8\%$ of the total delay),
and the signature verification (in the malicious setting that contributes an extra $2$--$6\%$ overhead).

\begin{figure}[!t] 
	\centering
			\hspace{-1em}
				\begin{subfigure}{.49\columnwidth}
%
%
\definecolor{A}{HTML}{e6194B}%
\definecolor{B}{HTML}{f58231}%
\definecolor{C}{HTML}{4363d8}%
\definecolor{D}{HTML}{911eb4}%
\definecolor{E}{HTML}{3cb44b}%
\definecolor{F}{rgb}{0.92900,0.69400,0.12500}%
\definecolor{G}{HTML}{808000}%
\definecolor{H}{HTML}{000000}%
\begin{tikzpicture}
\small	
	\begin{axis}[%
		width=0.74\textwidth,
		height=0.35\textwidth,
		at={(1.128in,0.894in)},
		scale only axis,
		xmin=0,
		xmax=6,
		xlabel={Number of assisting nodes},
		xtick={0, 1, 2, 3, 4, 5, 6},
		xticklabels={$ $, $3$, $ 7 $, $ 11 $, $ 15 $, $ 19 $},
            xlabel style={font=\footnotesize},
            ylabel style={font=\footnotesize},
            tick label style={font=\footnotesize},  
		ymin= 0,
		ymax= 1.8,
		ylabel = {Delay (ms)},
		ylabel shift=-5pt,
		yticklabel shift={0cm},
		axis background/.style={fill=white},
		legend columns=1,
		legend style={legend cell align=left, align=left, fill=none, draw=none,inner sep=-0pt, row sep=0pt,font=\tiny},
		legend pos = north west,
		%
		minor y tick num=5,
		]

       \addplot [color=A, solid, mark=diamond*, mark options={solid, A}]
		table[row sep=crcr]{%
                1 0.2989\\
			2 0.5621\\ 
			3 0.8351\\ 
			4 1.0962\\ 
			5 1.3599\\
		};
		\addlegendentry{\nano~(MS)}

        \addplot [color=E, solid, mark=asterisk, mark options={solid, E}]
		table[row sep=crcr]{%
                1 0.2386\\ 
			2 0.5025\\  
			3 0.7717\\ 
			4 1.0359\\ 
			5 1.2991\\
		};
		\addlegendentry{\nano~(SH)}

	\end{axis}

\end{tikzpicture}
					\caption{Setup delay}\label{subfig:exp:anc:delay}
			\end{subfigure}\hspace{2mm}
			\begin{subfigure}{.49\columnwidth}
%
%
\definecolor{A}{HTML}{e6194B}%
\definecolor{B}{HTML}{f58231}%
\definecolor{C}{HTML}{4363d8}%
\definecolor{D}{HTML}{911eb4}%
\definecolor{E}{HTML}{3cb44b}%
\definecolor{F}{rgb}{0.92900,0.69400,0.12500}%
\definecolor{G}{HTML}{808000}%
\definecolor{H}{HTML}{000000}%
\begin{tikzpicture}
\small	
	\begin{axis}[%
		width=0.74\textwidth,
		height=0.35\textwidth,
		at={(1.128in,0.894in)},
		scale only axis,
		xmin=0,
		xmax=6,
		xlabel={Number of assisting nodes},
		xtick={0, 1, 2, 3, 4, 5, 6},
		xticklabels={$ $, $3$, $ 7 $, $ 11 $, $ 15 $, $ 19 $},
            xlabel style={font=\footnotesize},
            ylabel style={font=\footnotesize},
            tick label style={font=\footnotesize},  
		ymin= 0,
		ymax= 19,
		ylabel shift=-5pt,
		yticklabel shift={0cm},
		axis background/.style={fill=white},
		legend columns=1,
		legend style={legend cell align=left, align=left, fill=none, draw=none,inner sep=-0pt, row sep=0pt, font=\tiny},
		legend pos = north west,
		%
		minor y tick num=5,
		]

       \addplot [color=A, solid, mark=diamond*, mark options={solid, A}]
		table[row sep=crcr]{%
                1 4.8049 \\ 
			2 7.8079\\  
			3 10.4518\\  
			4 13.6691\\ 
			5 16.4386\\ 
		};
		\addlegendentry{\nano~(MS)}

        \addplot [color=E, solid, mark=asterisk, mark options={solid, E}]
		table[row sep=crcr]{%
                1 2.1726\\ %
			2 4.6079\\  
			3 7.5499\\ 
			4 10.2491\\ 
			5 13.6237\\
		};
		\addlegendentry{\nano~(SH)}

	\end{axis}

\end{tikzpicture}%
					\caption{Aggregation delay}\label{fig:exp:anc:aggregation}
			\end{subfigure}

	\caption{User time in Setup and Aggregation phases}\label{fig:exp:ANC} 
\end{figure}

 \begin{figure}[!t] 
	\centering
			\hspace{-1em}
				\begin{subfigure}{.49\columnwidth}
				\begin{tikzpicture}
    \begin{axis}[
        width=0.8\textwidth,    
        height=0.35\textwidth,
        at={(1.128in,0.894in)},
        scale only axis,        
        xlabel={\#training rounds},
        xlabel style={font=\footnotesize},
        ylabel style={font=\footnotesize},
        tick label style={font=\footnotesize},        
        ylabel={Accuracy score},
        ylabel shift=-5pt,
        yticklabel shift={0cm},
        xmin=0, xmax=300,
        ymin=0, ymax=1,
        xtick={0,50,100,150,200,250,300},
        ytick={0,0.2,0.4,0.6,0.8,1},
        axis background/.style={fill=white},
        legend columns=1,
        legend style={legend cell align=left, align=left, fill=none, draw=none,inner sep=-0pt, row sep=0pt, font=\tiny},
        legend pos = south east,
        ymajorgrids=true,
        grid style=dashed,
    ]
    \addplot[color=blue, mark=none, thick, smooth] table {
        0 0.1328
        50 0.7527999999999999
        100 0.8484
        150 0.8624
        200 0.8752
        250 0.8832
        300 0.8896
    };
    \addlegendentry{FedAvg}

    \addplot[color=orange, mark=none, dashed, thick, smooth] table {
        0 0.1328
        50 0.7527999999999999
        100 0.8484
        150 0.8624
        200 0.8752
        250 0.8832
        300 0.8896
    };
    \addlegendentry{Flamingo}

    \addplot[color=purple, mark=none, densely dotted, thick, smooth] table {
        0 0.1328
        50 0.7527999999999999
        100 0.8484
        150 0.8624
        200 0.8752
        250 0.8832
        300 0.8896
    };
    \addlegendentry{\nano}

    \end{axis}
\end{tikzpicture}
					\vspace{-1.5em}
					\caption{Accuracy for EMNIST}\label{fig:exp:acc:emnist}
			\end{subfigure}\hspace{2mm}
			\begin{subfigure}{.49\columnwidth}
			\begin{tikzpicture}
    \begin{axis}[
        width=0.8\textwidth,
        height=0.35\textwidth,
        at={(1.128in,0.894in)},
        scale only axis,
        xlabel={\#training rounds},
        xlabel style={font=\footnotesize},
        ylabel style={font=\footnotesize},
        tick label style={font=\footnotesize},        
        ylabel shift=-5pt,
        yticklabel shift={0cm},
        xmin=0, xmax=300,
        ymin=0, ymax=1,
        xtick={0,50,100,150,200,250,300},
        ytick={0,0.2,0.4,0.6,0.8,1},
        axis background/.style={fill=white},
        legend columns=1,
        legend style={legend cell align=left, align=left, fill=none, draw=none,inner sep=-0pt, row sep=0pt, font=\tiny},
        legend pos = south east,
        ymajorgrids=true,
        grid style=dashed,
    ]
    \addplot[color=blue, mark=none, thick, smooth] table {
        0 0.18059701492537314
        50 0.6402985074626866
        100 0.7761194029850746
        150 0.8208955223880596
        200 0.8417910447761194
        250 0.853731343283582
        300 0.8597014925373134
    };
    \addlegendentry{FedAvg}
    
    \addplot[color=orange, mark=none, dashed, thick, smooth] table {
        0 0.18059701492537314
        50 0.6402985074626866
        100 0.7761194029850746
        150 0.8208955223880596
        200 0.8417910447761194
        250 0.853731343283582
        300 0.8597014925373134
    };
    \addlegendentry{Flamingo}

    \addplot[color=purple, mark=none, densely dotted, thick, smooth] table {
        0 0.18059701492537314
        50 0.6402985074626866
        100 0.7761194029850746
        150 0.8208955223880596
        200 0.8417910447761194
        250 0.853731343283582
        300 0.8597014925373134
    };
    \addlegendentry{\nano}

    \end{axis}
\end{tikzpicture}
					\vspace{-1.5em}
					\caption{{Accuracy for CIFAR-100}}\label{fig:exp:acc:cifar}
			\end{subfigure}

	\caption{Privacy-Preserving vs.~Plain FL Model Accuracy}\label{fig:exp:acc} 
\end{figure}

\myPara{Aggregation phase: Outbound bandwidth} 
Figure~\ref{fig:exp:BW} compares the outbound bandwidth costs of the Aggregation phase for all the entities in our protocol with the ones in MicroSecAgg and Flamingo.
We have observed that our protocol's outbound bandwidth cost for the server and user is considerably lower than those in the MicroSecAgg protocols and Flamingo.
As presented in Figure~\ref{fig:exp:comm:server}, the server in our semi-honest (resp.~malicious) protocol incurs $2,564\times$--$10,000\times$ (resp.~$2,564\times$--$10,000\times$) less outbound bandwidth overhead compared to MicroSecAgg protocols 
and $3\times$--$9\times$ (resp.~$3\times$--$9\times$) less outbound bandwidth overhead compared to Flamingo protocols.
Specifically, for $1000$ users, our protocol incurs $62,500$ KB (model size) in the semi-honest and $62,500$ KB plus an additional $64$ B (for signature) in the malicious setting, compared with $610,352$ MB in MicroSecAgg, $190,755$ KB in \FlamingoServerFirst and $575,520$ KB in \FlamingoServerSec.
In MicroSecAgg, the server needs to send the list of online users and signatures to all system users.
In Flamingo, the server notifies all regular users that the aggregation round is beginning and sends a list of regular online users and symmetric ciphertext to all decryptors.
Note that the overhead of distributing the final model to all users is captured in \nano, MicroSecAgg, and Flamingo.

As shown in Figure~\ref{fig:exp:comm:user}, the user outbound bandwidth cost in our semi-honest (resp.~malicious) protocol is $279\times$ (resp.~$278\times$) lower than the one in MicroSecAgg.
Additionally, it is $2\times$ (resp.~$2\times$) less than that of both regular users and decryptors in Flamingo.
Specifically,
the user outbound bandwidth is $60$ KB for the semi-honest and $63$ KB for the malicious protocol, while it is $17,439$~KB in MicroSecAgg.
In Flamingo, for a user range of $200$--$1000$, the outbound bandwidth is $130$--$133$ KB for regular users and $143$--$195$ KB for decryptors.
In addition, the user outbound bandwidth in \nano depends on the number of assisting nodes (which is a small constant), and thus, it does not increase as the number of users increases during the aggregation phase.
Specifically, the user transmits the masked model and the iteration number (along with the signature in the malicious setting) to the aggregation server and the iteration number (along with the signature in the malicious setting) to the three assisting nodes.

In MicroSecAgg, apart from sending the masked update, the user sends the aggregation of the shares for all online users to the server.
In Flamingo, regular users send masked updates to the server along with symmetric and asymmetric ciphertexts.
Decryptors resend the list of online regular users received from the server along with a signature to all other decryptors.
Decryptors also send the decrypted values requested by the server for unmasking.
The outbound bandwidth cost of the assisting nodes in \nano is small and \emph{independent} of the number of users.
Specifically, in the semi-honest setting, the node sends a message of size $64,008$ B consisting of the masking term ($64,000$ B), the iteration number ($4$ B) (for $t \leq 128$), and the length of the user list ($4$ B).
A signature of size $64$ B will also be transmitted for the malicious setting.

\subsubsection{Overall Computation Overhead}
We compare the overall computation overhead of {\nano}~with our counterparts in Figure~{\ref{fig:exp:anc:total}}.
The overall computation overhead includes the runtime of users, assisting nodes (decryptors for Flamingo), and the aggregation server combined.

As in Figure~{\ref{fig:exp:anc:total:setup}}, the overall setup computation overhead in {\nano} (for both semi-honest and malicious settings) is only around $150$ ms, compared to $2,885$ ms in MicroSecAgg$_1$ and $210$ ms in MicroSecAgg$_2$.
The overall setup time in Flamingo is significantly higher, reaching $36,720$ ms for Flamingo${64}$ and $302,490$ ms for Flamingo${128}$.

Figure~{\ref{fig:exp:anc:total:agg}} reports the overall computation overhead in the aggregation phase.
{\nano}~is up to $55241\times$ and $65264 \times$ faster than MicroSecAgg$_1$ and MicroSecAgg$_2$ and up to $41\times$ faster than Flamingo in malicious setting, respectively.

\subsubsection{User Overhead vs.~Number of Assisting Nodes}
In Figure~\ref{fig:exp:ANC}, we compare user computation delay across setup and aggregation phases for various assisting node ranges.
User computation only takes $0.24$--$1.3$ ms in the semi-honest setting (and $0.3$-$1.4$ ms in the malicious setting) to set up key materials for $3$ to $19$ assisting nodes (Figure~\ref{subfig:exp:anc:delay}).
The user computation time for the aggregation phase slightly increases with the number of assisting nodes, as shown in Figure~\ref{fig:exp:anc:aggregation}.
Specifically, for $3$--$19$ assisting nodes, the user computation time is $2.2$--$14$ ms in the semi-honest setting (and $4.8$--$16$ ms in the malicious setting) to compute the mask update (and signature in the malicious setting).

\subsubsection{Privacy-Preserving Model Training}
We adopt two TensorFlow FL datasets: 1) The Extended MNIST\mbox{~\cite{corr/abs-1812-01097}}
dataset containing $340,000$ training and $40,000$ test samples where each sample is a $28\times28 $ image in each of the $10$ classes.
We randomly assign around $220$ images sampled from all the classes to each user.
2) CIFAR-100 dataset\mbox{~\cite{tr/cifar09}}
containing $50,000$ training and $10,000$ test samples where each sample is a $32\times32$ image in each of the $100$ classes.
For CIFAR-100, we randomly assign around $100$ images sampled from all the classes to each user.
We train a multilayer perceptron for image classification for the EMNIST dataset and a CNN network with additional normalization layers for CIFAR-100.
This results in a weight vector containing $8,000\times 32$-bit elements for EMNIST and $500,000\times 32$-bit elements for CIFAR-100.
Figure{~\ref{fig:exp:acc}}
compares the accuracy of
\nano with\mbox{~\cite{sp/MaWAPR23}}
and our non-private baseline (FedAvg\mbox{~\cite{aistats/McMahanMRHA17}})
where the accuracy of the model is measured with a sparse categorical accuracy score in TensorFlow.
Compared to the non-private FedAvg, the final model's accuracy is not impacted by {\nano}.
Our experimental setup, including model selection and data distribution with balanced classes, mirrors that of our counterparts\mbox{~\cite{sp/MaWAPR23,aistats/McMahanMRHA17}}.
Hence, any changes to the model architecture or data distribution are expected to affect them equally.
For example, several studies on data heterogeneity in federated learning have demonstrated that imbalanced data can impact accuracy, which could identically affect the non-private FedAvg and the models trained with {\nano} or Flamingo.
 
\subsubsection{Performance of Proof of Aggregation Integrity}\label{sec:integrityperf}
We measure the performance of our technique based on $\APVC$ to achieve aggregation integrity against a malicious server.
In the setup phase,
the assisting node samples a random secret of size $32$ B, and sends its authenticated ciphertext ($56$ B) to all the users, resulting in a total outbound bandwidth of $11,200$--$56,000$ B for $200$--$1000$ users.
The encryption (or decryption on the user side) takes roughly $40$--$200$ ms for $200$--$1,000$ users.
In the aggregation phase, \emph{each} user computes the authenticated commitment with EC operations, which takes $2.6$ min (mostly due to unoptimized EC scalar multiplication).
However, optimization techniques such as Pippenger or low-level language implementation can reduce this computation time.
Additionally, it incurs $64$ B outbound bandwidth to send it to the server.
The server performs aggregation over the authenticated commitments received from all the users, which takes a total of $4.02$--$20.72$ ms (mostly due to EC point additions) and incurs a total of outbound bandwidth of $12,800$--$64,000$ B for $200$--$1000$ users.




\section{Conclusion and Future Work}\label{sec:conclusion}

Federated learning often involves highly distributed devices operating under limited bandwidth or power constraints.
Therefore, user dropouts are common, potentially disrupting the convergence of the final model or impacting its generalizability.
Existing secure aggregation protocols often require multiple communication rounds to successfully unmask the final model, with the dropout rate in each round increasing the communication overhead for all participants.

We propose \nano, an efficient secure aggregation protocol leveraging assisting nodes.
{\nano} takes only one round in the aggregation phase.
With this simple workflow,
the usual computation overhead dependent on the dropout rate can be minimized.
Also, the dropout rate could be lower since the online aggregation only takes one round.
\nano also provides proof of aggregation to the participating users.



One future work is to explore other privacy-preserving techniques, such as differential privacy (DP) and $k$-anonymity, to protect the intermediate and final model.
{\nano} design allows for the adoption of both instance-level \mbox{\cite{ccs/AbadiCGMMT016}} and client-level \mbox{\cite{corr/GeyerKN17}} DP methods to protect against inference attacks.
Moreover, leveraging real-world data will allow for a deeper investigation into these methods' computational and communication efficiency.

Our work did not consider protection against malicious users compromising the integrity of the final model.
Some existing approaches could be employed alongside our approach to fully protect the model's integrity
However, such methods would require examining user (private) data, necessitating verifiable computation techniques for processing encrypted data.
Doing so\mbox{~\cite{ccs/LiuXZ21}} often involves resource-intensive cryptographic methods, \eg, \mbox{~\cite{cacm/ParnoHG016}}.
Additionally, the absence of a ``ground truth'' becomes a significant challenge if malicious users have fabricated the data.
An alternative approach with different system assumptions, such as utilizing semi-trusted certifiers to inspect user private data, could offer an interesting direction for future research.

\section*{Acknowledgment}
The authors thank the reviewers and the shepherd for their insightful comments and constructive feedback.
The work of Rouzbeh Behnia was supported by the USF Office of Research (Sarasota-Manatee campus) through the Interdisciplinary Research Grant program. The work of Thang Hoang and Arman Riasi was supported in part by an unrestricted gift from Robert Bosch, 4-VA, and the Commonwealth Cyber Initiative (CCI), an investment in the advancement of cyber R\&D, innovation, and workforce development. For more information about CCI, visit \url{www.cyberinitiative.org}.

\bibliographystyle{IEEEtran}
\bibliography{ref}

\appendices

\section{Cryptographic Building Blocks}\label{sec:cryptobg}
A key exchange protocol allows two parties to agree securely on a symmetric key over a public channel.
 
\begin{definition}[Key Exchange]\label{def:keyexchange}
A key exchange protocol is a tuple of algorithms $\Sigma:(\kakeygen, \kaagree)$ defined as follows.
 
\begin{itemize}

\item $(\pk_\Sigma, \sk_\Sigma) \gets \kakeygen(1^\kappa)$:
Takes as input a security parameter $\kappa$ and outputs a key pair $(\pk_\Sigma, \sk_\Sigma)$.

\item $\x^{\mathtt{U}_2}_{\mathtt{U}_1} \gets \kaagree(\pk_\Sigma^{\mathtt{U}_1}, \sk_{\Sigma}^{\mathtt{U}_2})$: Takes as input the public key of any user $\pk_\Sigma^{\mathtt{U}_1}$ and the private key of user~$\mathtt{U}_2$, $\sk_{\Sigma}^{\mathtt{U}_2}$, and outputs a shared secret key $\x^{\mathtt{U}_2}_{\mathtt{U}_1}$.
\end{itemize}

\end{definition}
The security of such protocols requires 
any probabilistic polynomial time (PPT) adversary who is given two honestly generated public keys $\pk_\Sigma^{\user_1}$ and $\pk_\Sigma^{\user_2}$, 
but without either of the corresponding secret keys ($\sk_{\Sigma}^{\user_1}$ or $\sk_{\Sigma}^{\user_2}$), 
is unable to distinguish the shared secret $\x^{\user_1}_{\user_2}$ from a uniformly random value.

\begin{definition}[Digital Signatures]\label{def:sig}
A digital signature scheme for a messages space $\mathcal{M}_{\Pi}$ is defined by $\Pi$:$(\sgnkeygen$, $\sgnsign$, $\sgnverify)$:

\begin{itemize}

\item $(\pk, \sk) \gets \sgnkeygen(1^\kappa)$: Takes as input a security parameter $\kappa$ and outputs a key pair $(\pk_\Pi, \sk_\Pi)$.

\item $\sigma \gets \sgnsign(\sk_\Pi, m)$:
Takes as input $\sk_\Pi$ and a message $m \in \mathcal{M}_{\Pi}$ and outputs a signature $\sigma$.

\item $\{0, 1\} \gets \sgnverify(\pk_\Pi, m, \sigma)$: Takes as input $\pk_\Pi$, a message $m \in \mathcal{M}_{\Pi}$, a signature $\sigma$ and outputs $1$ if $\sigma$ is a valid signature under $\pk_\Pi$ for message $m$.
Otherwise, it outputs $0$.
\end{itemize}

Correctness requires that for any $m \in \mathcal{M}_{\Pi}$, $\sgnverify(\pk_\Pi, m, \sigma) = 1$, where $(\pk_\Pi, \sk_\Pi) \gets \sgnkeygen(1^\kappa)$ and $\sigma \gets \sgnsign(\sk_\Pi, m)$.
\end{definition}

Existential unforgeability against adaptive chosen message attacks is the de facto standard for signature schemes.

\begin{definition}[Authenticated encryption~\cite{asiacrypt/BellareN00}]
\label{def:ae} 
Given a shared secret key $\x \in \mathcal{K}_{\mathtt{AE}}$ in a key space $\mathcal{K}_{\mathtt{AE}}$, an authenticated encryption $\mathtt{AE}:(\mathtt{enc}, \mathtt{dec})$ is defined as follows.

\begin{itemize}
	\item $c \gets \mathtt{enc}(\x, m)$: Takes as input a shared secret key $\x$ and a message $m \in \mathcal{M}_{\mathtt{AE}}$ (for a message space $\mathcal{M}_{\mathtt{AE}}$) and outputs a ciphertext $c$ appended with its MAC.
	\item $\{m, \bot\} \gets \mathtt{dec}(\x, c)$: Takes as input a shared secret key $\x$ and a ciphertext $c$ and verifies the MAC
	and outputs either the original message $m$ or an error $\bot$.
\end{itemize}

\end{definition}
Indistinguishability against chosen-plaintext attack
requires that any PPT adversary has only a negligible advantage in distinguishing between ciphertexts of two different adversarially chosen messages.

\begin{definition}[Vector Commitment]
A vector commitment $\mathtt{VC}:(\Setup, \Comm)$ for a vector space $\set{M}^d_{\mathtt{VC}}$, randomness space $\set{R}_\mathtt{VC}$, and commitment space $\set{C}_\mathtt{VC}$ consists of the following PPT algorithms 
\begin{itemize}
	\item $\pp \gets \Setup(1^\kappa, d)$: Takes as input security parameter $\kappa$ and vector length $d$, and outputs public parameter $\pp$.

	\item $\cm \gets \Comm(\vec{x}, r)$: Takes as input a vector $\vec{x}$ and a randomness $r$, and outputs a committed value $\cm$.
\end{itemize}
\end{definition}

VC offers homomorphic property under $(\set{M}^d, *)$, $(\set{R}, *)$, $(\set{C}, \bullet)$ that
$\forall \vec{x}_1, \vec{x}_2 \in \set{M}^d$, $r_1, r_2 \in \set{R}$, 
$\VC.\Comm(\vec{x}_1, r_1) \bullet \VC.\Comm(\vec{x}_2, r_2) = \VC.\Comm(\vec{x}_1 * \vec{x_2}, r_1*r_2)$.

For security, we require binding and hiding:
\begin{itemize}
	\item For all integers $\secParam$ and PPT $\adv$, $\pp \gets \VC.\Setup(1^\secParam, d)$, 
	$(\vec{x}_0, \vec{x}_1) \gets \adv(\pp)$, 
 $\cm = \VC.\Comm(\vec{x}_b, r)$ where $b \getsRandom \{0, 1\}$, $r \getsRandom \set{R}$, and $b' \gets \adv(\pp, \cm)$, 
	there exists a negligible function $\negl$ such that
$\left|\Pr[b = b'] - \frac{1}{2}\right| \le \negl(\secParam)$

 \item For all integers $\secParam$ and PPT $\adv$, $\pp \gets \VC.\Setup(1^\secParam, d)$, and $(\vec{x}_0, \vec{x}_1, r_0, r_1) \gets \adv(\pp)$ where $\vec{x}_0, \vec{x}_1 \in \set{M}^n$, there exists a negligible function $\negl $ such that
		$\Pr[\VC.\Comm(\vec{x}_0, r_0) = \VC.\Comm(\vec{x}_1, r_1)] \le \negl(\secParam)$.
\end{itemize}

\begin{definition}\label{def:homoMAC:sec}
[Unforgeability modulo homomorphism]
Unforgeability of $\APVC$ is defined via a game $\mathsf{EUGame}_\adv(\kappa)$ between challenger $\mathcal{C}$ and adversary $\adv$:

\noindent \textbf{Setup.}
The challenger $\mathcal{C}$ samples a random secret $\rho \getsRandom \mathcal{K} $, and initializes an empty list $\mathcal{L}$ which records the responses to $\adv$'s queries in the form of $(\vec{x}, r)$ and can be queried s.t.~$\vec{x} \gets \set{L}(r)$.
 
\mySmallSkip
\noindent
\textbf{Query.} $\adv$ adaptively submits a query to $\mathcal{C}$ as $\big(\vec{x}, r\big)$, where $\vec{x}$ is a vector to be committed, and $r$ is randomly sampled.
		$\mathcal{C}$ rejects if $r \in \set{L}$.
		Otherwise, 
		$\mathcal{C}$ computes $\cm \gets \APVC.\Comm(\rho, \vec{x}, r)$ and sends $\cm$ to $\adv$.

\mySmallSkip
\noindent
\textbf{Output.}
$\adv$ outputs a vector $\vec{x}^*$, a commitment $\cm^*$, $m$ randoms $(r^*_1, \dots, r^*_m)$, and $m$ constants $(c^*_{1}, \dots, c^*_{m})$ such that 
 $\sum_{i = 1}^m c^*_i r^*_i \notin \mathcal{L}$, $\adv$\emph{wins} and the game outputs $1$ if 
 (i) 
 $\cm^* = \APVC.\Comm(\rho, \vec{x}^*, \sum_{i = 1}^m c^*_i r^*_i)$, 
	(ii) $(c^*_1, \dots, c^*_m) $ are not all zeros (trivial forgery), 
	and (iii) Unforgeability is w.r.t.~homomorphism: Either
		1) $\exists r^*_i \notin \set{L}$ for some $i \in [m]$, or 
		2 )$r^*_i \in \set{L}, \forall i \in [m]$ and $\vec{x}^* \ne \prod_{i = 1}^m (c_i \cdot \vec{x}_i) $, where $\vec{x}_i \gets \set{L}(r^*_i) $.
	$\APVC$ achieves \emph{unforgeability} if 
	$
	\Pr[\mathsf{EUGame}_\adv(\secParam)] \le \negl(\secParam).
	$
 \end{definition}

\section{Proof of Theorem 1}\label{prf:thm1}

\begin{proof}
Correctness directly follows from that of the underlying key exchange protocol.
Each user's update (\eg, $\mathbf{w_t^{\user_i}}$) is masked by the sum of $k$ masking terms computes as $\Vec{a}^{\user_i}_{t} = \sum_{j = 1}^k\PRF(\x^{\user_i}_{\node_j}, t)$, where the secure pseudorandom function $\PRF(\cdot)$ is invoked on the input of iteration number $t \in [T]$ and a shared secret key $\x^{\node_j}_{\user_i}$ between $\user_i$ and the $k$ assisting nodes $\node_j$ for $j \in \{1, \dots, k\}$.
The shared secret key $\x $ is derived in the Setup phase (Algorithm~\ref{alg:nanoSetup}) by invoking $\Sigma.\kaagree(\cdot)$ of the underlying key exchange protocol.
Following Algorithm~\ref{alg:nanoSetup}, $\user_i$ computes $\x^{\node_j}_{\user_i} \gets \Sigma.\kaagree(\sk^{\node_j}_{\Sigma}, \pk^{\user_i}_\Sigma) $ and $\node_j$ computes $\x^{\user_i}_{\node_j} \gets \Sigma.\kaagree(\sk^{\user_i}_{\Sigma}, \pk^{\node_j}_{\Sigma}) $, and the correctness of the underlying key exchange protocol ensures that $\x^{\node_j}_{\user_i} = \x^{\user_i}_{\node_j}$.
For the online user list $\Ulist_{\node_j, t}$, 
each assisting node computes $\Vec{a}^{\node_j}_{t} = \sum_{i = 1}^{|\mathcal{L}_{j, t}|}\PRF(\x^{\node_j}_{\user_i}, t)$.
Now, given $\x^{\node_j}_{\user_i} 
= \x^{\user_i}_{\node_j}$ for all $\user_i$ and $\node_j$, it is easy to see $\sum_{j = 1}^{k} \Vec{a}^{\node_j}_{t} = \sum_{i = 1}^{\Ulist_{\agg, t}}\Vec{a}^{\user_i}_{t}$ and
$
\sum_{i = 1}^{|\mathcal{L}_{\agg, t}|}\Vec{y}^{\user_i}_{t} - \sum_{j = 1}^k\Vec{a}^{\node_j}_t
= \sum_{i = 1}^{|\mathcal{L}_{\agg, t}|}\Vec{w}_t^{\user_i} +\sum_{i = 1}^{|\mathcal{L}_{\agg, t}|}\Vec{a}_t^{\user_i} - \sum_{j = 1}^k\Vec{a}^{\node_j}_t = \sum_{i = 1}^{|\mathcal{L}_{\agg, t}|}\Vec{w}_t^{\user_i}.$
 
This is satisfied when the number of participating honest parties $(1 - \delta)|\mathcal{P}_h| \geq \alpha|\mathcal{P}_h|$, where $\alpha$ is defined in Definition~\ref{def:alphasummation}.
\end{proof}

\section{Proof of Lemma 1}
\begin{proof}

Based on Definition~\ref{def:alphasummation}, the protocol requires the participation of at least $|\mathcal{P}_t|\geq \alpha|\mathcal{P}_h|$ participates for the final model to be computed.
However, note that the aggregation phase in \nano is one round.
More specifically, once the user sends their update and participation message in Algorithm~\ref{alg:nanoAgg}, Phase 1, Step 3, they can go offline and do not need to communicate any further with the aggregation server or the assisting nodes.
However, suppose one wants to consider the dropout of assisting nodes.
In that case, we can perform a simple secret sharing algorithm~\cite{cacm/Shamir79} and secret share the secret key of each assisting node among other assisting nodes in a $k$ out of $t$ manner.
Therefore, if one assisting node is offline, at least $k$ of the other assisting nodes need to be online to reconstruct its secret seed.
In this case, the protocol would require at least $k$ of the assisting nodes to be online in each iteration.
\end{proof}

\section{Proof of Theorem 2}\label{prf:thm2}
\begin{proof}
Following Definition~\ref{def:aggSec}, we construct a simulator $\Sim$ 
and prove via the standard hybrid argument, showing the indistinguishability between different hybrids.
We first define the behavior of our simulator $\Sim$ in each round of the Setup and Aggregation phase.
In the following, we let $\calP$ be the set of all parties, $\calP_h$ be honest parties, and $\calP_C$ be corrupt parties, \ie., $\calP_C = \calP \setminus \calP_h$.
We use the same notation for assisting nodes $\calA$.

\noindent
\emph{Setup phase:}
\begin{enumerate}
	\item Each honest user $\user_i$ and assisting node $\node_j$ follows the protocol in Algorithm~\ref{alg:nanoSetup}.

	\item For each corrupt user $\user_{i'} \in \calP_{C}$ and honest assisting node $\node_j \in \calA_H$, $\node_j$ computes and stores $\x_{\user_{i'}}^{\node_j} \gets \Sigma.\kaagree(\sk_\Sigma^{\node_j}, \pk_\Sigma^{\user_{i'}})$.
	\item For each honest user $\user_{i} \in \calP_H$ and corrupt assisting node $\node_{j'} \in \calA_C$, $\user_i$ computes and stores $\x^{\user_i}_{\node_j} \gets \Sigma.\kaagree(\sk_\Sigma^{\user_i}, \pk_\Sigma^{\node_{j'}})$.
	\item For each honest pair of user $\user_i$ and honest assisting node $\node_j$, $\Sim$ picks $\x_r \randasgn \mathcal{K}_\Sigma$ and sets $\x_{\user_i}^{\node_j} = \x^{\user_i}_{\node_j} = \x_r$.
\end{enumerate}

\noindent\emph{Aggregation phase:}

\begin{enumerate}
	\item In each iteration $t$ of the protocol, each honest user picks a random $\Vec{y'}_t^{\user_i}$ and sends $t$ to the assisting nodes and $(\Vec{y'}_t^{\user_i}, t)$ to the server.
	\item The aggregation server $\agg$ first adds the user $\user_i$ to the list $\Ulist_{\agg, t}$ and then calls the $\alpha$-summation ideal functionality $\mathcal{F}_{\Vec{x}, \alpha}(\Ulist_{\agg, t} \setminus \mathcal{P}_{C})$ (where $\mathcal{P}_{C}$ is the set of corrupt users) to get $\Vec{w}_t$.

\item Next, the simulator samples $\Vec{w'}_t^{\user_i} \randasgn \mathcal{M}^d$ for all $\user_i \in \Ulist_{\agg, t} \setminus \mathcal{P}_{C} $ such that $\Vec{w}_t = \sum_{i \in \mathcal{P}_{C}}\Vec{w'}_t^{\user_i} $ and computes $\Vec{a'}_{t}^{\user_i} = \Vec{y'}_t^{\user_i} - \Vec{w'}_t^{\user_i}$.
$\Sim$ sets 
$\Vec{a'}_{\node_j, t}^{\user_i}$
to be the random oracle output when queried 
$\x_{\node_j}^{\user_i}, t$ 
for each
$\node_j \in \calA_H$,
such that
$\Vec{a'}_{t}^{\user_i} = \sum_{\node_j \in \calA_H} \{\Vec{a'}_{\node_j, t}^{\user_i}\}$, and for each $\node_j \in \calA_H$, it computes $\Vec{a'}_{\user_i, t}^{\node_j} = \sum_{\user_i \in \calP_H}\Vec{a'}_{j, t}^{\user_i}$.
\end{enumerate}

We present our hybrids below.
The hybrids represent different views of the system as seen by corrupted entities, \eg, aggregation server, assisting nodes, and/or users), and the proof is based on the assumption that these views are computationally indistinguishable from each other.
The simulator $\Sim$ is used to construct these hybrids, and the adversary $\adv$ controls the corrupted entities.

\begin{description} 
\item [$\mathtt{Hyb0}$]
This is the real execution of the protocol where $\adv$ interacts with the honest entities.

\item [$\mathtt{Hyb1}$]
In this hybrid, a simulator that knows all the secrets of the honest parties (in each iteration) is introduced.
This is just a syntactic change.

\item [$\mathtt{Hyb2}$]
In this hybrid, we change the behavior of the simulated honest parties $\user_i \in \calP_H$ and assisting nodes $\node_j \in \calA_H$ by selecting a random shared secret key from the key space $\mathcal{K}_\mathtt{\Sigma}$ instead of instantiating the $\Sigma.\kaagree(\cdot)$ algorithm.

The security of the key exchange protocol (\eg, as in~\cite{ccs/BonawitzIKMMPRS17}) guarantees the indistinguishability of this hybrid with the one before.

\item [$\mathtt{Hyb3}$]
In this hybrid, each honest user $\user_i \in \calP_H$ replaces $\Vec{y}^{\user_i}_t$ it sends to the server with a random vector $\Vec{y'}^{\user_i}_t$.
As mentioned, we require at least one of the assisting nodes in our protocol to be honest.
Therefore, the indistinguishability of this hybrid is guaranteed since $\adv$ does not have knowledge on the honest entity's secret and hence in $\real$ the masked update will have the same distribution as $\Vec{y'}^{\user_i}_t$ in $\simul$.

\item [$\mathtt{Hyb4}$]
This hybrid replaces the aggregated masking term outputted by the honest assisting nodes $\node_i \in \calA_H$ with $\Vec{a'}_{\user_i, t}^{\node_j}$ (computed in Step 3 of the simulated Aggregation phase presented above) by calling the idea functionality $\mathcal{F}_{\Vec{x}, \alpha}(\Ulist_{\agg, t} \setminus \mathcal{P}_{C})$ and utilizing a random oracle.
Note that the view of this hybrid is indistinguishability with the previous hybrid since $\adv$ does not have any knowledge of the honest entities' shared secret and the distribution of the aggregate masking term in $\simul$ is identical to the one in $\real$.

\item [$\mathtt{Hyb5}$]
In this hybrid, $\Sim$ sends the output of the ideal functionality (as in Step 2 in the simulated Aggregation phase above) as the universal model update $\vec{w}_t$ for iteration $t$.
Note that the ideal functionality will not return $\bot$ based on the condition on the fraction of the honest users.
Hence, this hybrid is indistinguishability from the previous one, given the local updates are not known by $\adv$ in $\real$.

\end{description}
In the above, we have shown that the view of all the corrupted parties controlled by $\adv$ is computationally indistinguishable 
\end{proof}

\section{Proof of Theorem 3}\label{prf:thm3}
\begin{proof}
Following Definition~\ref{def:aggSec} and similar to the proof of Theorem~\ref{thm:semi-honest}, we prove the above theorem by the standard hybrid argument using the simulator $\Sim$.
In the following, we define the set of all parties as $\calP$, honest parties as $\calP_h$ and corrupt parties as $\calP_C$, \ie, $\calP_C = \calP \setminus \calP_h$.
We apply the same notation to the assisting nodes $\calA$.
 $\Sim$ behaves as follows in each round of the Setup and Aggregation phases.

\mySmallSkip
\noindent\emph{Setup phase:}
$\Sim$ works similar to the Setup phase in the proof of Theorem~\ref{thm:semi-honest}.

\mySmallSkip
\noindent
\emph{Aggregation phase:}

\begin{enumerate}
	\item In each iteration $t$ of the protocol, each honest user picks a random $\Vec{y'}_t^{\user_i}$.
	It then computes two signatures $\sigma^{\user_i}_{\agg} \gets \Pi.\sgnsign(\sk^{\user_i}_\Pi, m)$ and $\sigma^{\user_i}_{\node_j} \gets \Pi.\sgnsign(\sk^{\user_i}_\Pi, m')$ and sends $(m{, \sigma^{\user_i}_{\agg}})$ and $(m'{, \sigma^{\user_i}_{\node_j}})$ to the aggregation server and the assisting node $\node_j $ (for $j \in [1, \dots, k]$), respectively.
 \item $\Sim$ works similar to Steps~2 and~3 of the Aggregation phase in the proof of Theorem~\ref{thm:semi-honest}.
$\user_i$ to the list $\Ulist_{\agg, t}$ and then calls the $\alpha$-summation ideal functionality $\mathcal{F}_{\Vec{x}, \alpha}(\Ulist_{\agg, t} \setminus \mathcal{P}_{C})$ (where $\mathcal{P}_{C}$ is the set of corrupt users) to get $\Vec{w}_t$.
\end{enumerate}

We present our hybrids below and follow the same approach for transitioning between hybrids as we did in the proof of Theorem~\ref{thm:semi-honest}.

\begin{description} 
\item [$\mathtt{Hyb0}$]
This random variable is distributed identically to $\real$, \ie, the real execution of the protocol where $\adv$ interacts with the honest entities.

\item [$\mathtt{Hyb1}$]
In this hybrid, a simulator that has knowledge of all the secrets of the honest parties (in each iteration) is introduced.

The distribution of this hybrid is identical to the previous one.

\item [$\mathtt{Hyb2}$]
In this hybrid, we change the behavior of the simulated honest parties $\user_i \in \calP_H$ and assisting nodes $\node_j \in \calA_H$ by selecting a random shared secret key from the key space $\mathcal{K}_\mathtt{\Sigma}$ instead of instantiating the $\Sigma.\kaagree(\cdot)$ algorithm.

The indistinguishability of this hybrid with the one before is guaranteed by the security of the instantiated key exchange protocol.
For instance, with the Diffie-Hellman key exchange protocol, this is guaranteed by the 2ODH assumption~\cite{ccs/BonawitzIKMMPRS17}.

\item [$\mathtt{Hyb3}$]
In this hybrid, each honest user $\user_i \in \calP_H$ samples a random vector $\Vec{y'}^{\user_i}_t$ for the update domain.
It also computes a signature $\sigma^{\user_i}_{\agg} \gets \Pi.\sgnsign(\sk^{\user_i}_\Pi, <\Vec{y'}^{\user_i}_t, t>)$ and sends $(\Vec{y'}^{\user_i}_t, \sigma^{\user_i}_{\agg})$ to the server.
Firstly, the indistinguishability of $\sigma^{\user_i}_{\agg}$ is guaranteed by the underlying signature scheme.
Secondly, as mentioned, we require at least one of the assisting nodes in our protocol to be honest.
Therefore, the indistinguishability of this hybrid is guaranteed since $\adv$ does not have knowledge on the honest entity's secret and hence in $\real$ the masked update will have the same distribution as $\Vec{y'}^{\user_i}_t$ in $\simul$.

\item [$\mathtt{Hyb4}$]
In this hybrid, we call the idea functionality $\mathcal{F}_{\Vec{x}, \alpha}(\Ulist_{\agg, t} \setminus \mathcal{P}_{C})$ and utilize random oracles to replace the aggregated masking term outputted by the honest assisting nodes $\node_i \in \calA_H$ with
$\Vec{a'}_{\user_i, t}^{\node_j}$ (computed in Step 3 of the simulated Aggregation phase presented above).
The output of this hybrid is $( \Vec{a'}_{\user_i, t}^{\node_j}, \sigma^{\node_i}_{\agg})$ where $\sigma^{\node_i}_{\agg} \gets \sgnsign(\sk^{\node_j}_\Pi, m'')$.
The indistinguishability of $\sigma^{\node_i}_{\agg}$ is provided via the security of the underlying signature scheme.
Hence, the view of this hybrid is indistinguishability with the previous hybrid since $\adv$ does not have any knowledge of the honest entities' shared secret and the distribution of the aggregate masking term in $\simul$ is identical to the one in $\real$.

\item [$\mathtt{Hyb5}$]
In this hybrid, $\Sim$ sends the output of the ideal functionality (as in Step 2 in the simulated Aggregation phase above) as the universal model update $\vec{w}_t$ for iteration $t$.
Note that the ideal functionality will not return $\bot$ based on the condition on the fraction of the honest users.
Hence, this hybrid is indistinguishability from the previous one, given the local updates are not known by $\adv$ in $\real$.

\end{description}
In the above, we have shown that the view of all the corrupted parties controlled by $\adv$ is computationally indistinguishable 
\end{proof}

\section{Proof of Lemma 2}\label{prf:lemma2}

\begin{proof}


	To prove this lemma, we first prove the local model's privacy in the presence of $\APVC$, following the proof of Theorems~\ref{thm:semi-honest} and~\ref{thm:malicious} and how it affects the certain hybrids in the proofs.
	We then prove the aggregation integrity.
	Theorems~\ref{thm:semi-honest} and~\ref{thm:malicious}.
	As highlighted in the protocol description, in its plain form, we require the initiating assisting node $\node_j$ to be honest.
	This proof applies to the other extensions discussed in the paper.

	In the Setup phase and $\mathtt{Hyb2}$ of the proofs $\node_j$ samples random $\rho$ and $c_{\node_j}^{\user_i}$.
	Note that the $indistinguishability$ of this hybrid would also rely on the security of the underlying vector commitment scheme $\APVC$ and the authenticated symmetric encryption scheme $\mathtt{AE}$.
	In $\mathtt{Hyb3}$, the user will also output $\cm^{\user_i}_t$.
	Note that the $indistinguishability$ of this hybrid relies on the \emph{hiding} property of the underlying vector commitment scheme $\APVC$.

	In $\mathtt{Hyb5}$, $\Sim$ sets $x \gets \Vec{g}^{\Vec{w}_t\cdot\rho}$ and sends $x$ along with the final model $\Vec{w}_t$.
	Therefore, the view of all the corrupted parties controlled by $\adv$ is computationally indistinguishable.

	Next, we prove the integrity.
	The unforgeability of $\APVC$ permits the user to verify whether the linear combination of local weight models (which is the aggregation) has been computed correctly.
	Let $\vec{w}'_t = \vec{w}_t + \epsilon$ be the tampered aggregated model, where $\vec{\epsilon}$ is the error vector introduced by the malicious server and $\vec{w}_t = \sum_{i = 1}^n \vec{w}^{\user_i} $ is the aggregated model that is supposed to be computed honestly in the iteration $t$.
	For the user to accept $\vec{w}'_t$, the server needs to somehow generate a valid $\vec{g}^{\vec{w}'_t \cdot \rho} = \vec{g}^{(\sum_{i = 1}^n{\vec{w}_t^{\user_i}} + \vec{\epsilon}) \cdot \rho} $ from $\{\vec{g}^{ \vec{w}^{\user_i} \cdot \rho} h^{r^{\user_i}}\}_{i \in [n]}$.
	Given that $\rho$ and all $\{r^{\user_i}\}_{i \in [n]}$ are secret to the server, this happens with only a negligible probability due to the unforgeability (by Definition~\ref{def:homoMAC:sec}) of $\APVC$, which only permits the server to perform the predefined linear combination of $\vec{g}^{ \vec{w}^{\user_i} \cdot \rho }$, but nothing else.
\end{proof}
\end{document}